\documentclass[5p,]{elsarticle}

\usepackage{cite}
\usepackage{graphicx}
\usepackage{amsmath, amssymb}
\usepackage{url}
\usepackage{booktabs}
\usepackage{multirow}
\usepackage{hyperref}
\usepackage{xcolor}
\usepackage{booktabs}

\usepackage{soul}
\sethlcolor{yellow}

\usepackage{color}

\journal{Information Sciences}

\begin{document}

\begin{frontmatter}

 \title{Ideological Isolation in Online Social Networks: A Survey of Computational Definitions, Metrics, and Mitigation Strategies}

\author[1,2]{Xiaodan Wang}

\author[2]{Yanbin Liu}

\author[3]{Shiqing Wu}

\author[2]{Ziying Zhao}

\author[5]{Yuxuan Hu}

\author[2]{Weihua Li\corref{cor1}}

\author[4]{Quan Bai\corref{cor2}}

\cortext[cor1]{Corresponding author: weihua.li@aut.ac.nz}
\cortext[cor2]{Corresponding author: quan.bai@utas.edu.au}

\affiliation[1]{organization={School of Engineering, Yanbian University},
                city={Yanji},
                postcode={133002},
                state={Jilin Province},
                country={China}}
                
\affiliation[2]{organization={Department of Data Science and Artificial Intelligence, Auckland University of Technology},
                city={Auckland},
                postcode={1010},
                country={New Zealand}}

\affiliation[3]{organization={Faculty of Data Science, City University of Macau},
                city={Macau SAR},
                country={China}}

\affiliation[4]{organization={School of Information and Communication Technology, 
                               University of Tasmania},
                city={Hobart},
                postcode={7001},
                state={Tasmania},
                country={Australia}}
\affiliation[5]{organization={Integrated Marine Observing System, 
                               University of Tasmania},
                city={Hobart},
                postcode={7004},
                state={Tasmania},
                country={Australia}}

\begin{abstract}
The proliferation of online social networks has significantly reshaped the way individuals access and engage with information. While these platforms offer unprecedented connectivity, they may foster environments where users are increasingly exposed to homogeneous content and like-minded interactions. Such dynamics are associated with selective exposure and the emergence of filter bubbles, echo chambers, tunnel vision, and polarization, which together can contribute to ideological isolation and raise concerns about information diversity and public discourse.
This survey provides a comprehensive computational review of existing studies that define, analyze, quantify, and mitigate ideological isolation in online social networks. We examine the mechanisms underlying content personalization, user behavior patterns, and network structures that reinforce content-exposure concentration and narrowing dynamics. This paper also systematically reviews methodological approaches for detecting and measuring these isolation-related phenomena, covering network-, content-, and behavior-based metrics. We further organize computational mitigation strategies, including network-topological interventions and recommendation-level controls, and discuss their trade-offs and deployment considerations. By integrating definitions, metrics, and interventions across structural/topological, content-based, interactional, and cognitive isolation, this survey provides a unified computational framework. It serves as a reference for understanding and addressing the key challenges and opportunities in promoting information diversity and reducing ideological fragmentation in the digital age. 

\end{abstract}


\end{frontmatter}

\section{Introduction}
With the rapid evolution of social platforms, 
such as Facebook, TikTok, WeChat, X (formerly Twitter), and YouTube, algorithmic curation and social connectivity have become primary gateways to information. 
Many people rely on these social media platforms to stay informed, learn, be entertained, and engage in civic life. This convenience comes with a concern: highly personalized feeds and socially clustered interaction patterns can narrow the breadth of information and fragment public discourse \citep{de2022modelling}. Users may increasingly encounter ideologically aligned content and like-minded peers, while rarely engaging with dissenting views. 
This condition, referred to as \textit{ideological isolation} \citep{hu2024influence}, manifests through related phenomena including filter bubbles \citep{flaxman2016filter, areeb2023filter}, echo chambers \citep{mahmoudi2024echo,terren2021echo}, tunnel vision \citep{elaad2022tunnel}, exposure bias \citep{jin2023survey}, and polarization \citep{flamino2023political}, which jointly shape how beliefs form, reinforce, and spread online.

Ideological isolation poses several intertwined challenges. 
First, information overload and selection make it difficult to maintain balanced exposure in the presence of ranking, recommendation, and limited attention \citep{levy2019echo, valensise2023drivers, metzler2024social}. 
Second, network structure (e.g., modular communities, homophily) shapes the pathways and speed of cross-cutting information flow, often reinforcing within-group interaction \citep{tornberg2022digital}. 
Third, algorithmic filtering and amplification can magnify alignment signals derived from past behavior, creating reinforcing feedback loops \citep{valensise2023drivers, de2022modelling}. 
Fourth, cognitive and social dynamics, including selective exposure, confirmation bias, and the false-consensus effect, encourage users to favor attitude-consistent content and avoid contrary perspectives \citep{metzler2024social}. 
Finally, external amplifiers, such as coordinated campaigns or bots, can distort exposure diversity \citep{yan2023exposure, wang2023evidence, li2024social}. Addressing these challenges requires an interdisciplinary approach that connects computational modeling, measurement, and model design \citep{liao2024modeling}. 

\begin{figure*}
    \centering
    \includegraphics[width=0.9\linewidth]{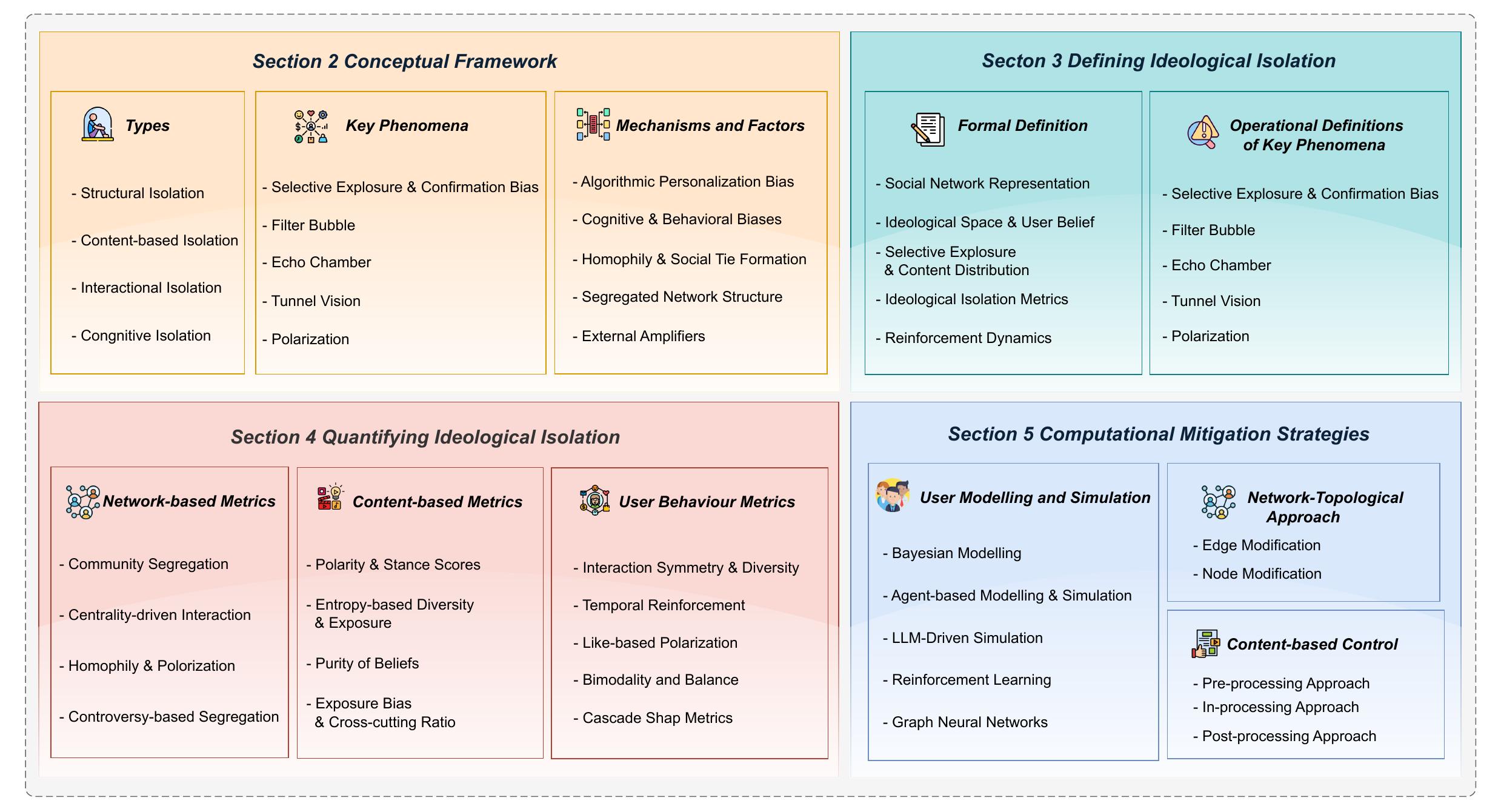}
    \caption{Overview of ideological isolation in online social networks. The figure illustrates the structure of this survey, spanning conceptual framing, formal definition, quantitative measurement, and computational mitigation.} 
    \label{fig:overall}
\end{figure*}

Over the past decade, these issues have motivated a growing body of research \citep{rodrigues2024social} across computer science, information science, and the social sciences. However, existing work remains fragmented, with inconsistent definitions, heterogeneous measurements, and mitigation strategies scattered across graph-based approaches and recommendation system adjustments. In this survey, we provide a methodological synthesis from a computational perspective to study ideological isolation. We organize the literature around (i) a compact conceptual framework that distinguishes four interrelated isolation types; (ii) formal definitions that place users and content in a shared ideological space and specify visibility, exposure, and reinforcement dynamics; (iii) operational metrics spanning network, content, and behavior indicators; and (iv) computational mitigation strategies at both the network topology and recommendation algorithm levels. 

To provide an end-to-end view of ideological isolation, we structure the paper as shown in Figure~\ref{fig:overall}:
\begin{enumerate}
  \item \textbf{Conceptual Framework (Section~\ref{sec:conceptual_framework})}. 
  We introduce four types of ideological isolation: structural, content-based, interactional, and cognitive. 
  Then, we describe the key phenomena associated with these types, including selective exposure \& confirmation bias, filter bubbles, echo chambers, tunnel vision, and polarization. 
  Finally, we analyze the mechanisms and factors that drive these phenomena, including algorithmic personalization, cognitive and behavioral biases, homophily and tie formation, network segregation, and external amplifiers.
  Figure~\ref{fig:overall} summarizes these relationships. 

  \item \textbf{Definitions and Problem Formulations (Section \ref{sec:def_ideological_isolation})}. 
  We formalize the social network structure, ideological space \& user beliefs, together with content visibility and exposure mechanisms, ideological isolation indices, and belief reinforcement dynamics.
  Building on these formal foundations, we provide operational definitions and precise conditions for key phenomena associated with ideological isolation, including selective exposure \& confirmation bias, filter bubbles, echo chambers, tunnel vision, and polarization.

  \item \textbf{Quantifying Ideological Isolation (Section~\ref{sec:quantifying_ideological_solation})}. 
  We review network-based metrics (e.g., segregation, homophily, and controversy-oriented measures), content-based metrics (e.g., polarity, diversity, and exposure-related measures), and behavior-based metrics (e.g., interaction, temporal, and cascade-based measures), and harmonize notation to support comparability across methods.

  \item \textbf{Mitigation Strategies (Section~\ref{sec:mitigation_strategies})}. 
  We review computational mitigation strategies, including user modeling and simulation (e.g., agent-based and learning-based models), network-topological interventions (e.g., edge and node modification), and content-based control mechanisms (e.g., pre-, in-, and post-processing).
  We discuss trade-offs between accuracy, utility, and exposure diversity.

  \item \textbf{Research Challenges and Directions (Section~\ref{sec:challenges_direction})}. 
  We outline key open problems, including cross-platform measurement, causal identification under exposure bias, human-in-the-loop evaluation, robustness to adversarial amplification, and practical deployment constraints for diversity-aware systems.
\end{enumerate}

{\textbf{Differences from Existing Surveys.}}
Existing surveys largely treat these themes in isolation, focusing on recommender-system phenomena such as echo chambers \citep{hartmann2025systematic}, filter-bubble dynamics \citep{areeb2023filter}, and exposure bias \citep{jin2023survey}, with no integrative computational framework that connects formal definitions, metrics, and mitigation strategies. 
In contrast, this survey contributes (i) a concept-to-operations bridge that maps isolation types to concrete metrics and intervention strategies;  
(ii) a unified formalism for exposure, diversity, and reinforcement that aligns network, content, and behavior dimensions; and (iii) a mitigation taxonomy that spans both network topology and recommendation pipelines with explicit measurement targets.

{\textbf{Main Contributions.}} The contributions of this survey are summarized as follows:
\begin{itemize}
  \item We propose a compact conceptual framework of ideological isolation that distinguishes structural, content-based, interactional, and cognitive types and relates them to driving mechanisms and external amplifiers.
  
  \item We provide unified formal definitions of content visibility, exposure centroids and variance, and reinforcement dynamics, and specify operational conditions for ideological isolation phenomena, including selective exposure, filter bubbles, echo chambers, tunnel vision, and polarization. 
  
  \item We systematically organize the quantification methods across network-, content-, and behavior-based metrics with harmonized notation and guidance for interpretation and normalization.
  
  \item We organize computational mitigation strategies at the network-topological and recommendation-system levels, analyzing trade-offs and deployment considerations for increasing cross-cutting exposure and semantic diversity.
  
\end{itemize}

\section{Conceptual Framework}
\label{sec:conceptual_framework}

This section presents a unifying framework for studying ideological isolation that integrates key phenomena, types, and mechanisms and clarifies their relationships. We begin with the phenomena observed in practice, organize them into types of ideological isolation that provide a coherent taxonomy for analysis, and then examine the mechanisms and factors that give rise to these types and drive the observed phenomena.

\subsection{Key Phenomena}

\paragraph{\textbf{Selective exposure and confirmation bias}} Individuals tend to prefer belief-congruent information; on modern platforms, however, this tendency is co-produced with feed design. 
Large-scale evidence from Facebook shows that user choices and ranking algorithms reduce cross-cutting exposure, yielding ideologically aligned consumption even when diverse items are available \citep{bakshy2015exposure}. 
 At the same time, the recommender-systems literature emphasizes that exposure (what is shown) is distinct from preference (what is liked or clicked), and that feedback loops in exposure can entrench visibility imbalances and perceived consensus, an agenda now formalized under fairness-aware recommendation \citep{jin2023survey}. Moving beyond descriptive correlations, recent causal-inference approaches explicitly model item exposure and user satisfaction as separate processes to debias learning from logs, helping disentangle human confirmation tendencies from system-induced selection effects \citep{liao2024modeling}.

\paragraph{\textbf{Filter Bubble}} Filter bubbles are commonly described as an exposure environment shaped by personalization in which the breadth of viewpoints presented to a user narrows over time \citep{burbach2019bubble}. Empirically, web-scale tracking shows that platform pathways (e.g., social referrals and search) can increase the ideological distance of news exposure, indicating that bubbles are primarily an exposure-level phenomenon rather than merely a matter of individual preference \citep{flaxman2016filter}. 
At the same time, the recommender-systems literature cautions that evidence is mixed across platforms and metrics: systematic reviews report both indications of narrowing exposure and counter-examples where recommendation pipelines maintain diversity, underscoring the need to distinguish what is shown from what is preferred and to measure diversity explicitly \citep{areeb2023filter}. 
Finally, user-interface controls that allow users to adjust system curation can increase awareness of bubble effects and reduce extremity, although their impact on political diversity remains mixed \citep{liu2024does}.

\paragraph{\textbf{Echo Chamber}} Echo chambers are generally understood as social contexts characterized by dense connections among ideologically like-minded users, where content exposure is largely confined within the community, leading to high source concentration and limited cross-community exchange. Comparative reviews show that this phenomenon is typically examined through the network lens (e.g., homophily and community structure) and interaction patterns, although evidence for its prevalence and effects varies across platforms and measurement choices~\citep{terren2021echo}. 
Recent syntheses map the field’s core dimensions, attributes, mechanisms, modeling and detection approaches, and metrics, and highlight the absence of a single canonical definition, underscoring the need to specify how echo chambers are operationalized in a given study \citep{choi2020rumor}. 
A recent systematic review attributes disagreements about existence and impact to divergent conceptualizations and operationalizations (e.g., exposure, topology, or attitudes), showing that empirical conclusions depend strongly on the chosen measurement strategy and context \citep{hartmann2025systematic}.

\paragraph{\textbf{Tunnel Vision}} Tunnel vision in social networks refers to the aspect-level narrowing of online discourse, where collective attention converges on a limited subset of an issue’s facets while alternative angles and context are sidelined, reducing informational diversity \citep{hewitt2001beyond}. Unlike filter bubbles or echo chambers, which emphasize the sources of content and ideological alignment, tunnel vision emphasizes what is being discussed, i.e., a content-level concentration of frames or aspects, even in the absence of explicit community segregation.

\paragraph{\textbf{Polarization}} Polarization denotes the separation of audiences or communities into distinct, often opposing, camps with limited middle ground. 
In practice, polarization manifests as multi-modal distributions of opinions and stances on controversial topics, as well as fragmented communities with sparse bridging ties. 
Distributional analyses of controversies on social media reveal clear splits into discernible camps, providing a content-level signature of polarization \citep{qiu2019investigating}. Unlike the other phenomena, polarization is an emergent, system-level outcome, typically characterized at the population or network scale \citep{iandoli2021impact}.

\subsection{Types of Ideological Isolation}

Following the overview of key phenomena in Section 2.1, this subsection introduces four interrelated types of ideological isolation: structural, content-based, interactional, and cognitive. These types are not mutually exclusive and provide a conceptual basis for organizing observed phenomena and connecting measurement and mitigation in later sections. 

\paragraph{\textbf{Structural Isolation}} Users reside in segregated or highly modular clusters in the social network. 
This type is commonly associated with \textit{echo chamber effects} and is often reinforced by \textit{algorithmic curation} that preferentially links users with similar preferences. 
AI-driven personalization can further intensify this form of isolation by shaping interaction pathways through invisible structures.
    
 \paragraph{\textbf{Content-based Isolation}} 
 Users are consistently exposed to ideologically skewed, repetitive, or redundant content. 
 This type of isolation is closely related to \textit{filter bubbles}, and the cognitive narrowing described by \textit{tunnel vision}. Both algorithmic ranking and user preferences reinforce such exposure patterns.
    
 \paragraph{\textbf{Interactional Isolation}}  
 User interactions are predominantly homophilic, with ties formed mainly among like-minded individuals and limited engagement with dissenting views. This form of isolation is linked to \textit{selective exposure}, \textit{confirmation bias}, and \textit{echo chamber formation}, where network structure and user behavior jointly reduce ideological diversity in interactions.
    
 \paragraph{\textbf{Cognitive Isolation}} 
 Users exhibit reduced openness to opposing perspectives, reinforcement of prior beliefs, and susceptibility to the false consensus effect. 
 This form of isolation reflects a deepening of \textit{confirmation bias} and often manifests as \textit{tunnel vision}, arising from sustained exposure to aligned viewpoints and limited engagement with alternative opinions.

\subsection{Mechanisms and Contributing Factors of Ideological Isolation}

Having outlined the key phenomena and types of ideological isolation, we now examine the mechanisms and contributing factors that generate and reinforce these patterns in online social networks. 

\paragraph{\textbf{Algorithmic Personalization Bias}} AI-powered recommender systems selectively curate content based on prior engagement \citep{wu2023soac}, inevitably reinforcing filter bubbles and reducing content diversity \citep{hu2024influence,areeb2023filter,wang2025balancing}. Algorithmic bias and cognitive bias are key contributors to excessive personalization in online social networks, particularly in recommender systems. 
Algorithmic bias is introduced during system design and implementation, while cognitive bias (e.g., confirmation bias) influences user interactions and feedback loops \citep{guess2023social}. 
Although fairness and diversity have received increasing attention in recommender-system research, most systems still emphasize personalization, thereby contributing to ideological isolation \citep{areeb2023filter,jin2023survey}. 
Simulation studies show that algorithmic curation and selective exposure can fragment public discourse into ideological ``tunnels'' \citep{de2022modelling}. 
Model-to-data analyses further indicate that interactions between algorithmic curation and user selectivity can reproduce and amplify polarized and fragmented information landscapes \citep{valensise2023drivers}.

\paragraph{\textbf{Cognitive and Behavioral Biases}} Users are inclined toward selective exposure and confirmation bias, seeking attitude-consistent information while avoiding opposing views \citep{knobloch2015confirmation}. Echo chambers can restrict users to homogeneous content, driven jointly by network homophily and structure, algorithmic curation, and selective engagement \citep{hartmann2025systematic}, thereby fostering biases such as the false consensus effect (FCE), in which individuals overestimate how much others share their views \citep{luzsa2019links}. 
Luzsa and Mayr show that higher FCE is associated with prolonged social media use, homogeneous networks, and lower ambiguity tolerance \citep{luzsa2021false}. Additionally, individuals may adopt group-conforming opinions either to reduce cognitive dissonance or to gain social acceptance \citep{iandoli2021impact}, a phenomenon that has also been encoded in opinion-aware influence maximization models \citep{dai2022opinion}. 
The lack of dissenting views in users’ feeds reduces critical scrutiny and encourages further ideological rigidity \citep{luzsa2021false}.

\paragraph{\textbf{Homophily and Social Tie Formation}} Individuals tend to form connections with others who share similar beliefs, leading to interactional isolation and the formation of echo chambers. 
Simulation models demonstrate that users may sever ties with disagreeing peers, thereby reinforcing homophily \citep{kashima2021ideology}. 
This behavior can be incentivized in polarized environments, where adopting provocative or oversimplified messages attracts greater attention and followers \citep{iandoli2021impact, peng2023role}. 
As like-minded agents cluster within echo chambers, attempts to shift user preferences often fail due to entrenched social structures \citep{von2024integrating}. 

\paragraph{\textbf{Segregated Network Structure}} Community clusters with high modularity restrict cross-group information flow, exacerbating topological isolation \citep{newman2006modularity}. Ferraz de Arruda et al. proposed a framework that simulates how algorithmic filtering and homophilic interactions amplify opinion polarization. They validated the model on real Twitter networks, demonstrating that such systems fragment discourse into ideologically segregated clusters \citep{de2022modelling}. These network structures limit exposure to diverse viewpoints and reinforce ideological rigidity.

\paragraph{\textbf{External Amplifiers}} Social bots, coordinated campaigns, and misinformation introduce artificial signals and distort information diversity, deepening ideological silos. 
In particular, coordinated information tactics, such as bots or sponsored content, can exacerbate selective exposure and algorithmic bias, making it harder for users to access diverse, high-quality information \citep{shao2018spread, huszar2022algorithmic, duan2025botdmm}. 
These amplifiers operate synergistically with platform algorithms and social dynamics to distort public discourse, creating environments in which ideologically isolated communities can thrive. 
In addition to these mechanisms, some studies examine content creation and reception as dual filtering pathways that drive tunnel vision \citep{andok2023religious}. 
Content is shaped by subjective values, organizational rules, and external incentives, such as market or political pressures. 
Meanwhile, users may actively avoid content they did not select, further fragmenting shared discourse and undermining common ground \citep{andok2023religious}. 

\begin{figure*}
    \centering
    \includegraphics[width=1\linewidth]{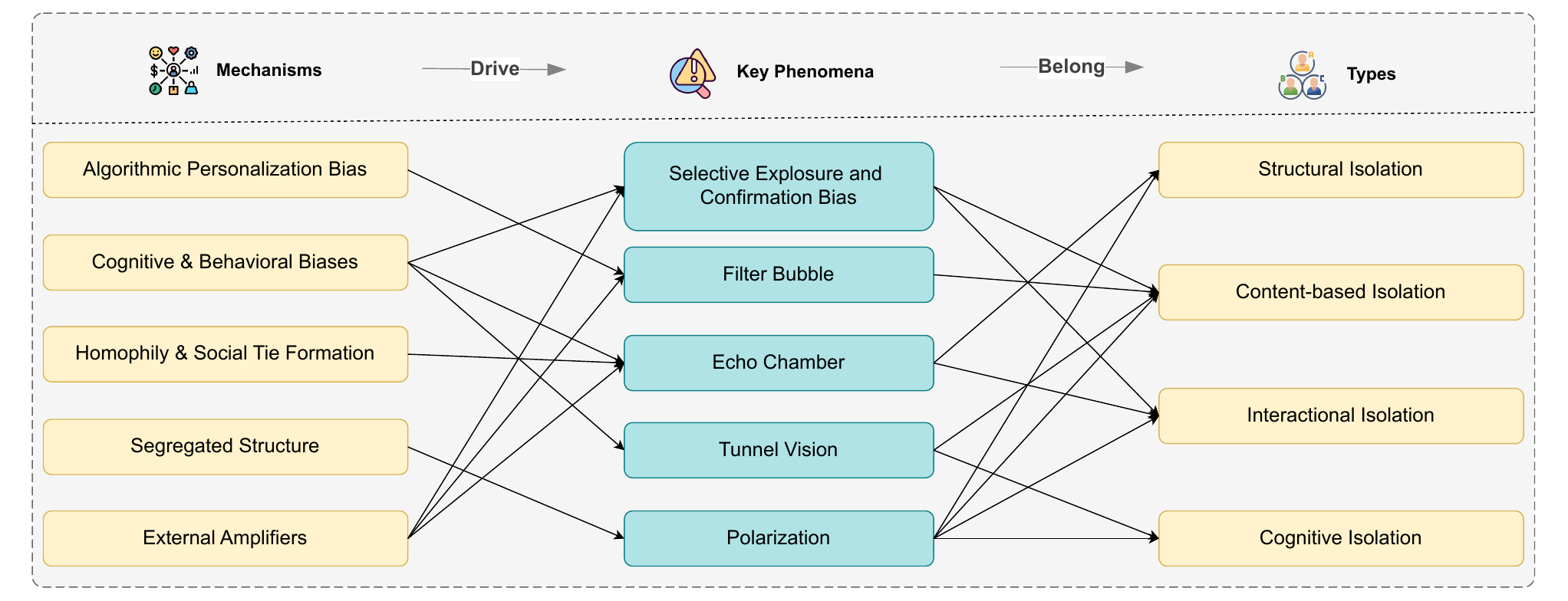}
    \caption[Mechanisms Phenomena Types map of ideological isolation]{Mechanisms Phenomena Types map of ideological isolation. 
    This figure maps five mechanisms, algorithmic personalization bias, cognitive \& behavioral biases, homophily \& tie formation, segregated network structure, and external amplifiers, to key ideological isolation phenomena, including selective exposure and confirmation bias, filter bubbles echo chamber, tunnel vision, and polarization, and further organizes these phenomena into four types: structural, content-based, interactional, and cognitive.}

    \label{fig:Mechanisms_Phenomena_Types}
\end{figure*}

In summary, this section establishes a conceptual framework for describing ideological isolation through observable phenomena and organizing it into four interrelated types. As illustrated in Figure~\ref{fig:Mechanisms_Phenomena_Types}, upstream mechanisms, including algorithmic personalization, cognitive and behavioral biases, homophily and social tie formation, segregated network structures, and external amplifiers, give rise to selective exposure and confirmation bias, filter bubbles, echo chambers, tunnel vision, and polarization. These phenomena, in turn, correspond to structural, content-based, interactional, and cognitive forms of ideological isolation. This layered mapping from mechanisms to phenomena to types underpins the remainder of the survey.

\section{Defining Ideological Isolation}
\label{sec:def_ideological_isolation}
This section formalizes ideological isolation as a computational construct. We define the networked setting, the ideological space for beliefs and content, and the exposure process that generates user feeds. On this basis, we introduce an isolation index that combines belief–exposure alignment with exposure diversity, along with a reinforcement dynamic for belief updating under repeated exposure. We then state operational criteria for five phenomena: selective exposure, filter bubbles, echo chambers, tunnel vision, and polarization, enabling consistent diagnosis across datasets and platforms.
\subsection{Formal Definition}

In this subsection, we present formal definitions of ideological isolation, focusing on its structural and dynamic properties in online social networks. We introduce key concepts, including user belief representations, content exposure profiles, and network interaction patterns, along with their associated notations and metrics. These formal elements provide a unified lens for capturing related phenomena such as filter bubbles, echo chambers, and tunnel vision, enabling quantitative analysis of how algorithmic filtering and social interactions shape the evolution of ideological landscapes.

\subsubsection{Social Network Representation}

Let the online social network be represented as a directed graph:
\[
G = (U, E),
\]
where $U = \{u_1, u_2, \dots, u_N\}$ is the set of users, and $E \subseteq U \times U$ is the set of directed edges representing social relationships, such as follower/followee, friend, and subscription \citep{lu2011link}.

For each user \(u_i \in U\), we define \(\varGamma(u_i) = \{ u_j \in U \mid (u_j, u_i) \in E \}\) as the set of in-neighbors, i.e., users whose content is visible to \(u_i\). Depending on the platform’s structure, this may reflect incoming edges, e.g., followees on X or friends whose posts appear in WeChat Moments. This neighborhood plays a central role in shaping user exposure on platforms that do not rely heavily on algorithmic recommendations, such as WeChat, where the content feed is largely provided by \(\varGamma(u_i)\), occasionally interleaved with advertisements or sponsored content\citep{li2020friend}.

Meanwhile, each user $u_i$ maintains a content feed $\mathcal{F}_i(t)$ at time $t$, which is shaped by the underlying logic of the platform. On many platforms, such as TikTok and YouTube, this feed is curated by recommender systems that leverage prior engagement history, social proximity, and behavioral signals to prioritize content\citep{hosseinmardi2024causally,vombatkere2024tiktok}. However, on other platforms, such as WeChat, $\mathcal{F}_i(t)$ primarily consists of content posted by the user’s social connections (e.g., friends), with limited algorithmic intervention apart from occasional advertisements\citep{zhang2018mobile}. In such cases, the structure of the user’s social network itself plays a dominant role in shaping ideological exposure, reinforcing the effects of homophily and social filtering over algorithmic personalization.

\subsubsection{Ideological Space and User Beliefs}

We define a continuous ideological space $\mathcal{I} \subset \mathbb{R}^d$ to represent the latent ideological dimensions along which both users and content can be situated. Each user $u_i \in U$ is associated with an ideological belief vector $\mathbf{b}_i \in \mathcal{I}$, which reflects their political orientation, social values, or other issue-based stances \citep{li2022unsupervised}. The dimension $d$ can vary depending on the granularity of ideological representation. For example, $d=1$ may represent a left-right political spectrum, while $d>1$ allows modeling multiple orthogonal ideologies (e.g., economic, social, environmental) \citep{li2020friend}. Similarly, let \(\mathcal{C} = \{c_1, c_2, \dots, c_M\}\) denote the global set of content items posted on the platform. Each item \(c_j \in \mathcal{C}\) may represent a post, comment, video, or article. At any time \(t\), each user \(u_i\) is exposed to a subset of these content items, forming their personalized content feed \(\mathcal{F}_i(t) \subseteq \mathcal{C}\).

Each content item $c_j \in \mathcal{C}$ is embedded in the same ideological space and is represented by a vector $\mathbf{v}_j \in \mathcal{I}$. These embeddings can be derived from textual content, metadata, creator attributes, or user engagement patterns using machine learning models, such as topic modeling, sentiment-aware embeddings, or supervised classifiers trained on ideological annotations \citep{monti2021learning}.
To formalize ideological alignment, we define a content-user affinity function $\phi: \mathcal{I} \times \mathcal{I} \rightarrow \mathbb{R}$ that measures the proximity or relevance of content to a user's beliefs. A simple way is the negative Euclidean distance:
\begin{equation}  
\phi(\mathbf{b}_i, \mathbf{v}_j) = -\lVert\
\mathbf{b}_i - \mathbf{v}_j\lVert \,,
\end{equation}  
where $\lVert \cdot \rVert$ denotes the Euclidean norm, so that ideologically closer content receives higher affinity scores. Alternatively, cosine similarity or kernel-based functions can be used to capture nonlinear ideological closeness.

We define a content visibility function \( P(c_j \mid u_i, t) \), which denotes the probability that a content item \( c_j \in \mathcal{C} \) appears in the feed \( \mathcal{F}_i(t) \) of user \( u_i \) at time \( t \). This probability is shaped by multiple interacting forces, including user-specific preferences, social connectivity, and platform-level mechanisms. For algorithmically curated platforms, \( P(c_j \mid u_i, t) \) is often determined by a ranking function that incorporates engagement signals, prior interactions, ideological affinity \( \phi(\mathbf{b}_i, \mathbf{v}_j) \), and temporal relevance. On socially driven platforms, e.g., WeChat, the visibility is constrained by the network topology, such that \( P(c_j \mid u_i, t) > 0 \) primarily if \( c_j \) is authored by some neighbor \( u_j \in \varGamma(u_i) \), where \( \mathcal{C}_j \subseteq \mathcal{C} \) denotes content posted by user \( u_j \) \citep{li2020friend}. Moreover, platform-imposed constraints, such as attention limits, freshness filtering, or injected advertisements, further restrict \( \mathcal{F}_i(t) \subseteq \mathcal{C} \), resulting in a highly selective exposure process. These visibility dynamics provide the foundation for understanding ideological alignment, belief reinforcement, and exposure narrowing in subsequent sections.

\subsubsection{Selective Exposure and Content Distribution}

Recall that given \(\mathcal{F}_i(t) \subseteq \mathcal{C}\), then $\mathcal{F}_i(t) = \{c_1, \dots, c_k\}$ denotes the set of content shown to user $u_i$ at time $t$. The exposure distribution of $u_i$ is:
\[
P_i(t) = \{ p_j(t) \,|\, p_j(t) = \Pr[c_j \in \mathcal{F}_i(t)] \} \,,
\]
following prior work that formalizes exposure as a probability distribution over feed content in social platforms \citep{bakshy2015exposure}.

The ideological exposure profile of user $u_i$ is defined as:
\[
\mu_i(t) = \sum_{c_j \in \mathcal{F}_i(t)} p_j(t) \cdot \mathbf{v}_j \,,
\] 
which represents the expected ideological position of the content user $u_i$ is exposed to at time $t$, analogous to continuous exposure modeling used in recent computational studies of information spread and recommender influence \citep{mehrotra2018towards,bovet2019influence}.

These formalizations provide a foundation for understanding selective exposure, belief reinforcement, and narrowing of ideological exposure in subsequent sections.

\subsubsection{Ideological Isolation Index}

This formal expression defines a conceptual isolation index that integrates belief-exposure alignment and content diversity. It provides the theoretical foundation for empirical measurement approaches \citep{barbera2015birds}.

\begin{equation}
\text{II}_i(t) = \| \mu_i(t) - \mathbf{b}_i \| + \alpha \cdot \text{Var}_{c_j \in \mathcal{F}_i(t)}(\mathbf{v}_j),
\label{eq:ideological_isolation}
\end{equation}
here \( \mu_i(t) \) is the ideological exposure profile of user \( u_i \), defined as the expected ideological position of content in their feed; \( \mathbf{b}_i \) is the user’s ideological belief vector; \( \left\| \mu_i(t) - \mathbf{b}_i \right\| \) measures the ideological proximity between the user’s beliefs and the average ideology of the content they are exposed to, smaller values indicate higher alignment and potential reinforcement of beliefs; \( \text{Var}_{c_j \in \mathcal{F}_i(t)}(\mathbf{v}_j) \) denotes the variance of the ideological positions of the content items in the feed, quantifying the degree of ideological diversity, specifically:
\[
\text{Var}_{c_j \in \mathcal{F}_i(t)}(\mathbf{v}_j) = \frac{1}{|\mathcal{F}_i(t)|} \sum_{c_j \in \mathcal{F}_i(t)} \left\| \mathbf{v}_j - \mu_i(t) \right\|^2,
\]
where \( \alpha > 0 \) is a tunable parameter that controls the weight assigned to content diversity.

Thus, the first term \( \left\| \mu_i(t) - \mathbf{b}_i \right\| \) captures ideological reinforcement, while the second term \( \alpha \cdot \text{Var}_{c_j \in \mathcal{F}_i(t)}(\mathbf{v}_j) \) penalizes ideological narrowness. A user is considered more ideologically isolated when they are exposed to content closely aligned with their own beliefs and lacking ideological variance, i.e., when both terms are small.

\subsubsection{Reinforcement Dynamics}
\label{subsubsec:reinforcement-dynamic}

The evolution of a user's ideological beliefs can be modeled as a gradual adjustment influenced by the content they are exposed to. Specifically, we define the belief update function for user \( u_i \) at time \( t \) as:
\[
\mathbf{b}_i(t+1) = \mathbf{b}_i(t) + \eta \cdot \left( \mu_i(t) - \mathbf{b}_i(t) \right), 
\]
where \( \mathbf{b}_i(t) \in \mathbb{R}^d \) denotes the ideological belief vector of user \( u_i \) at time \( t \); \( \mu_i(t) \in \mathbb{R}^d \) refers to the ideological centroid of the content feed \( \mathcal{F}_i(t) \), representing the average ideological position of content the user sees; \( \eta \in [0,1] \) is the susceptibility parameter, which controls the extent to which a user updates their beliefs in response to their exposure \citep{jiang2024opinion}.

This can be viewed as a weighted interpolation between the user's current belief and the ideological signal of the content they consume. When \( \eta = 0 \), the user is completely resistant to change, and their belief remains static. When \( \eta = 1 \), the user fully adopts the content exposure average.

Over time, if the content a user is exposed to is consistently aligned with their current beliefs (i.e., \( \mu_i(t) \approx \mathbf{b}_i(t) \)), the update will be small in magnitude but will reinforce the same belief direction, leading to ideological entrenchment. Conversely, if the exposure is diverse or ideologically distant, the user’s beliefs may shift, depending on the value of \( \eta \).

Importantly, when the variance of the ideological content in \( \mathcal{F}_i(t) \) is low (i.e., content is homogeneous), the directional pull of \( \mu_i(t) \) becomes stable and focused. As a result, repeated updates using this mechanism will steadily reinforce the user's current ideological stance and reduce openness to opposing views. This process captures the essence of reinforcement dynamics observed in echo chambers and filter bubbles.

The summary of notations is given in Table \ref{tab:notation}. 
\begin{table*}[!htbp]
\centering
\caption{Summary of Notations}
\label{tab:notation}
\begin{tabular}{ll}
\toprule[1pt]
\textbf{Symbol} & \textbf{Description} \\
\midrule
\(G = (U, E)\) & Online social network represented as a directed graph with users \(U\) and edges \(E\). \\
\(U\) & Set of users: \(\{u_1, u_2, \dots, u_N\}\). \\
\(E \subseteq U \times U\) & Directed edges denoting social relationships (e.g., follower/followee, friends). \\
\(\varGamma(u_i)\) & In-neighbors of user \(u_i\), i.e., users whose content is visible to \(u_i\). \\
\(\mathcal{F}_i(t) \subseteq \mathcal{C}\) & Personalized content feed of user \(u_i\) at time \(t\). \\
\(\mathcal{I} \subseteq \mathbb{R}^d\) & Ideological space shared by users and content. \\
\(\mathbf{b}_i \in \mathbb{R}^d\) & Ideological belief vector of user \(u_i\). \\
\(\mathcal{C}\) & Set of all content items (posts, comments, videos, etc.). \\
\(\mathcal{C}_j \subseteq \mathcal{C}\) & Content authored or reposted by user \(u_j\). \\
\(c_j\) & A single content item. \\
\(\mathcal{C}_a \subseteq \mathcal{C}\) & Aspect-specific content subset, i.e., item \(c_j\) that pertain to aspect \(a\). \\
\(\mathbf{v}_j \in \mathbb{R}^d\) & Ideological embedding of content item \(c_j\). \\
\(\phi(\mathbf{b}_i, \mathbf{v}_j)\) & Affinity score between user belief and content, e.g., \(-\|\mathbf{b}_i - \mathbf{v}_j\|\). \\
\(P(c_j \mid u_i, t)\) & Probability of content \(c_j\) appearing in \(u_i\)'s feed at time \(t\). \\
\(P_i(t)\) & Exposure distribution over content in \(\mathcal{F}_i(t)\). \\
\(\mu_i(t)\) & Ideological exposure profile of user \(u_i\); expectation over content vectors in their feed. \\
\(\text{Var}_{c_j \in \mathcal{F}_i(t)}(\mathbf{v}_j)\) & Variance of ideological embeddings of content in the feed. \\
\(\text{II}_i(t)\) & Ideological isolation score of user \(u_i\) at time \(t\). \\
\(\alpha > 0\) & Weighting parameter controlling the importance of content diversity in the isolation metric. \\
\(\text{Polarization}(G)\) & Network-level polarization index based on ideological distance between connected users. \\
\(\mathbb{I}[u_j \in \varGamma(u_i)]\) & Indicator function: 1 if \(u_j\) is a neighbor of \(u_i\), 0 otherwise. \\
\(\eta \in [0,1]\) & Susceptibility parameter controlling belief update strength. \\
\(\mathbf{b}_i(t+1)\) & Updated ideological belief of user \(u_i\) after exposure at time \(t\). \\
\bottomrule[1pt]
\end{tabular}
\end{table*}

\subsection{Operational Definitions of Key Phenomena}

Based on the formal definitions presented earlier, we generally characterize the existence of key ideological phenomena in online social networks, including selective exposure, filter bubbles, echo chambers, tunnel vision, and polarization. These problem definitions provide a unified structural and belief-based perspective, although alternative formulations may be employed in specific settings, depending on platform mechanics, user behavior, or analytical goals.

\subsubsection{Selective Exposure and Confirmation Bias}

Selective exposure refers to the tendency of individuals to seek out information that aligns with their existing beliefs, often driven by confirmation bias. Large-scale studies indicate that content exposure and recommendations are typically ideologically aligned with users’ prior beliefs, reflecting selective exposure and confirmation bias \citep{bakshy2015exposure}.

\begin{equation}
\mathbb{E}_{c_j \in \mathcal{F}_i(t)} [\phi(\mathbf{b}_i, \mathbf{v}_j)] \gg 0 \,.
\end{equation}

Alternatively, using the ideological exposure profile $\mu_i(t)$, the condition becomes:

\begin{equation}
\left\| \mu_i(t) - \mathbf{b}_i \right\| \leq \epsilon,
\end{equation}

\noindent where $\mu_i(t)$ is the expected ideological position of content in user $u_i$'s feed, and $\epsilon > 0$ is a small threshold indicating strong alignment with user belief $\mathbf{b}_i$.

\subsubsection{Filter Bubble}

Filter bubbles arise when algorithmic systems prioritize ideologically aligned content, thus limiting diversity \citep{ahmmad2025trap}. This occurs when content visibility is determined primarily by ideological similarity:

\begin{equation}
P(c_j \mid u_i, t) \propto \phi(\mathbf{b}_i, \mathbf{v}_j),
\end{equation}

\noindent and the ideological variance of exposed content is low:

\begin{equation}
\text{Var}_{c_j \in \mathcal{F}_i(t)}(\mathbf{v}_j) \leq \delta,
\end{equation}

\noindent where $\delta > 0$ is a small threshold for ideological diversity.

\subsubsection{Echo Chamber}
An echo chamber refers to an environment in which a user’s opinions, ideological stance, or beliefs are reinforced through repeated interactions with neighbors or information sources that share similar tendencies and attitudes \citep{hartmann2025systematic}.

\begin{align}
\forall u_j \in \varGamma(u_i), \quad \left\| \mathbf{b}_j - \mathbf{b}_i \right\| \leq \epsilon,\qquad |\Gamma(u_i)| \ge \gamma \,, \\
c_j \in \mathcal{F}_i(t) \Rightarrow c_j \in \mathcal{C}_k \text{ for some } u_k \in \varGamma(u_i) \,.
\end{align}
Here, all neighbours \(u_j\) of user \(u_i\) (within the visible set \(\Gamma(u_i)\)) 
hold similar ideological beliefs \(b_j\) within a small tolerance \(\epsilon\), 
and the neighbourhood size exceeds a minimum threshold \(\gamma\); consequently,The content displayed in the user’s feed \(F_i(t)\) primarily comes from users within the same neighbourhood \(\Gamma(u_i)\) \citep{cinelli2020echo}.

\subsubsection{Tunnel Vision}

Tunnel vision is a discourse-level concentration of attention in which prolonged selective exposure, amplified by filter bubbles and echo chambers, leads users to focus predominantly on a single aspect of an event while neglecting alternatives. 

Recall that $\mathcal{F}_i(t)\subseteq \mathcal{C}$ denotes the feed shown to user $u_i$ at time $t$, with exposure probabilities $p_j(t)=\Pr[c_j\in \mathcal{F}_i(t)]$. For each aspect $a$, define the subset
\[
\mathcal{C}_a=\{\,c_j\in \mathcal{C}:\ \text{item $c_j$ pertains to aspect }a\,\}.
\]

We say that user $u_i$ exhibits tunnel vision at time $t$ when the largest aspect share of exposure exceeds $1-\gamma$:
\begin{equation}
\label{eq:tunnel-vision-minimal}
\max_{a}\ \sum_{c_j\in F_i(t)\cap C_a} p_j(t)\ \ge\ 1-\gamma,
\end{equation}
for some small $\gamma>0$. Equivalently, there exists an aspect $a^*$ such that items about $a^*$ account for at least a $(1-\gamma)$ fraction of the user’s exposure. This minimal criterion captures aspect concentration succinctly and can be complemented by entropy-based measures when a continuous score is desired.

\subsubsection{Polarization}

Polarization is a macro-level phenomenon in which ideological distances increase across the network. It is present when the average pairwise ideological distance between connected users is high \citep{hohmann2023quantifying}:

\begin{equation}
\text{Polarization}(G) = \frac{1}{|U|^2} \sum_{i,j} \left\| \mathbf{b}_i - \mathbf{b}_j \right\| \cdot \mathbb{I}[u_j \in \varGamma(u_i)],
\end{equation}

\noindent where \(\left\| \mathbf{b}_i - \mathbf{b}_j \right\|\) represents the belief distance between user \(u_i\) and user \(u_j\), typically computed using Euclidean distance. $\mathbb{I}$ is an indicator function and $\varGamma(u_i)$ is the set of visible or interacting neighbors of $u_i$, specifically: 
\[
\mathbb{I}[u_j \in \varGamma(u_i)] =
\begin{cases}
1, & \text{if } u_j \in \varGamma(u_i)  \\
0, & \text{otherwise}
\end{cases}
\]

Polarization exists if 
\begin{equation}
\text{Polarization}(G) \geq \epsilon,
\end{equation}
and the social network exhibits community fragmentation, with high modularity driven by ideological clustering \citep{nair2024cross}.

\section{Quantifying Ideological Isolation}
\label{sec:quantifying_ideological_solation}

\begin{figure*}
    \centering
    \includegraphics[width=1\linewidth]{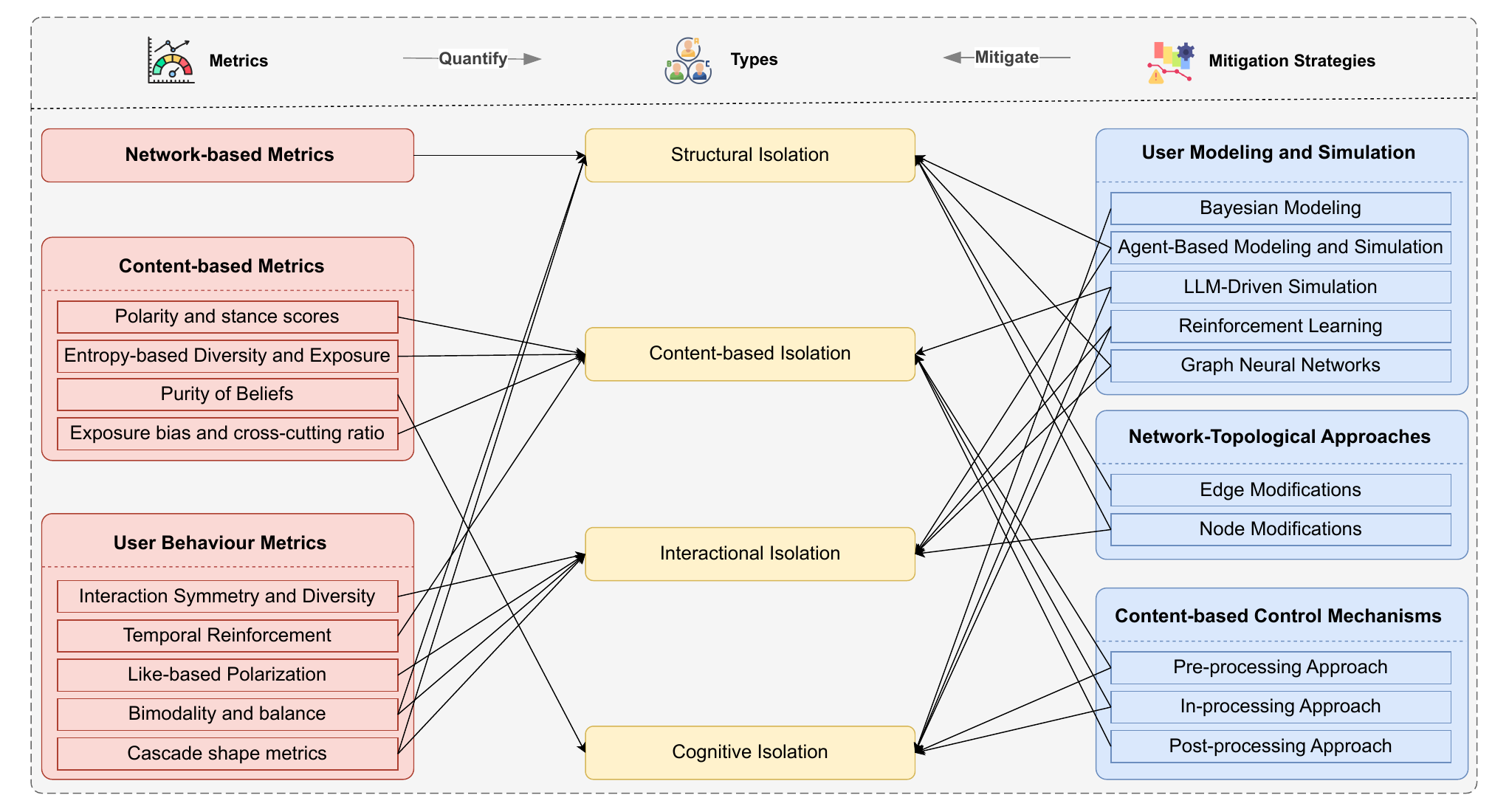}
    \caption[Metrics–Types–Mitigations map of ideological isolation]{Metrics–Types–Mitigations map of ideological isolation. This figure aligns three metric families (network-based, content-based, user behavior) with four isolation types (structural, content-based, interactional, cognitive) and indicates how mitigation strategies (user modeling \& simulation, content-based controls, network-topological approaches) act on the diagnosed type.}

    \label{fig:Metrics–Types–Mitigations}
\end{figure*}

In this section, we present a comprehensive set of quantitative approaches to assess ideological isolation in online social networks.  
It opens with Figure~\ref{fig:Metrics–Types–Mitigations} as an orienting key: network-, content-, and user-behavior metrics diagnose structural, content-based, interactional, and cognitive isolation, respectively, and guide the mitigation families introduced in Section 5.
Together, they provide a multifaceted understanding of how isolation emerges, persists, and manifests at both individual and community levels. All the symbols used for this section are listed in Table \ref{tab:symbols_quantification}. 

\begin{table}[!htbp]
\caption{Key symbols used in Section \ref{sec:quantifying_ideological_solation}}
\label{tab:symbols_quantification}
\centering
\small
\begin{tabular}{p{2cm} p{6cm}}
\toprule[1pt]
\textbf{Symbol} & \textbf{Description} \\
\midrule
$Q$ & Modularity of a partition. \\
$\Phi(S)$ & Conductance of subset $S \subset U$. \\
$\lambda_2(\mathcal{L})$ & Algebraic connectivity (Fiedler value) of Laplacian $\mathcal{L}$. \\
$\mathcal{L}$ & (Unnormalized) graph Laplacian. \\
$EI$ & External-Internal index of homophily. \\
$r_{\mathrm{attr}}$ & Categorical assortativity (ideology/camp mixing on edges). \\
$DL(M)$ & Map Equation (Infomap) description length for partition $M$. \\
$C_D(u_i)$ & Degree centrality of node $u_i$. \\
$C_E(u_i)$ & Eigenvector centrality of node $u_i$. \\
$C_C(u_i)$ & Closeness centrality of node $u_i$. \\
$d(i,j)$ & Shortest-path distance between nodes $u_i$ and $u_j$. \\
$C_B(u_i)$ & Betweenness centrality of node $u_i$. \\
$\mathrm{RWC}$ & Random Walk Controversy score on two-way partition. \\
$P_{ab}$ & Transition prob.\ for walk from $a$ to $b$ in $\{X,Y\}$. \\
$\mathrm{BC}$ & Boundary connectivity between partitions. \\
$\Delta_{ij}$ & Opinion distance $|p_i - p_j|$ between users $u_i$ and $u_j$. \\
$H_{\text{topic}}$ & Topic entropy. \\
$\mathrm{Spread}$ & Avg.\ pairwise cosine distance. \\
$\mathrm{Purity}$ & Share of users whose inferred ideology matches the community majority. \\
$\mathcal{C}$ & User set of a community. \\
$\mathbb{I}[\cdot]$ & Indicator function used in Purity. \\
$H$ & Entropy of a user’s exposure distribution. \\
$\mathbf{p}^{(u_j)}$ & Normalized exposure distribution for user $u_j$. \\
$H_{u_j}$ & User-level exposure entropy. \\
$\mathrm{EB}_i$ & Exposure bias of user $u_i$. \\
$\mathbf{v}_i \in \mathbb{R}^d$ & Embedding vector (user/content) in $d$-dimensional space. \\
$\mathrm{ES}_u$ & Engagement Symmetry for user $u$. \\
$\mathrm{IE}_u(t)$ & Interaction entropy at time $t$ for user $u$ (over $p_i^u(t)$). \\
$d_t$ & Euclidean gap between successive recommendations. \\
$D_{\mathrm{KL}}(P_t \Vert P_{t+1})$ & KL divergence between exposure distributions at $t$ and $t\!+\!1$. \\
$H(P_t)$ & Entropy of exposure distribution at time $t$. \\
$\rho(u)$ & Like-based polarization score for user $u$. \\
$\mathrm{BC}(t)$ & Bimodality coefficient at time $t$. \\
$\mathrm{BR}(t)$ & Balance ratio at time $t$. \\
$SV$ & Structural virality.  \\
\bottomrule[1pt]
\end{tabular}
\end{table}

\subsection{Network-based Metrics}
\label{subsec:network-metrics}

\subsubsection{Community segregation}
Community segregation occurs when a social network fragments into densely interconnected groups that share few links with the rest of the graph. Because most interactions, information flows, and social reinforcement occur within these clusters, individuals are repeatedly exposed to like-minded peers while seldom encountering discordant viewpoints. Such structural insulation amplifies confirmation bias and is therefore a key mechanism through which \emph{ideological isolation}, the tendency to hold increasingly homogeneous opinions, emerges at the individual level.

The strength of segregation is often quantified by \emph{modularity}~$Q$ and related indices (e.g., the $M$-value), together with conductance and algebraic connectivity~\citep{kratzke2023find}. Modularity is defined as
\begin{equation}
Q \;=\; \frac{1}{2m}\sum_{i,j} 
\Bigl(A_{ij} - \frac{k_i k_j}{2m}\Bigr)\,
\delta(c_i,c_j) \,,
\label{eq:modularity}
\end{equation}
where $A_{ij}$ is the edge weight, $k_i$ is the degree of node~$i$, $c_i$ is its community assignment, and $m$ is the total number of edges. A high $Q$ indicates that the network can be partitioned into well-separated communities, a signature of echo chambers. 

\emph{Conductance} offers a complementary view by measuring how well-separated a subset of nodes is from the rest of the graph \citep{zhang2018understanding}. For a set of nodes $S \subset U$, conductance is defined as
\[
\Phi(S) = \frac{|\partial S|}{\min(\mathrm{vol}(S), \mathrm{vol}(\bar{S}))} \,,
\]
where $\partial S$ is the set of edges with one endpoint in $S$ and one in $\bar{S} = U \setminus S$, and $\mathrm{vol}(S) = \sum_{i \in S} k_i$ is the total degree (volume) of nodes in $S$. A lower $\Phi(S)$ indicates a more segregated community.

Another important metric is the \emph{algebraic connectivity}, where $D$ is the diagonal degree matrix and $A$ the adjacency matrix \citep{fiedler1973algebraic}. Smaller $\lambda_{2}(L)$ indicates weaker global connectivity and the presence of structural bottlenecks—i.e., the graph admits good cuts and can contain well-separated clusters.
For degree-heterogeneous networks, the \emph{normalized} Laplacian
\[
\hat{L} \;=\; I - D^{-1/2} A D^{-1/2}
\]
provides a scale-invariant alternative; its Fiedler value $\lambda_{2}(\hat{L})$ plays an analogous role and is linked to cut quality via Cheeger-type bounds. While $\lambda_{2}$ itself is not ideology-specific, ideological isolation can be assessed spectrally by evaluating the Rayleigh quotient of the ideology partition with $\hat{L}$,
\[
\mathcal{R}(x) \;=\; \frac{x^\top \hat{L}\, x}{x^\top x} \,,
\]
which is equivalent to the normalized cut/conductance of that split; smaller values then indicate sparser cross-ideology connectivity.

These measures provide a multifaceted view of the network’s community structure and its potential to foster ideological isolation.

\subsubsection{Homophily and polarization}
Homophily refers to the social phenomenon where individuals are more likely to form connections with others who share similar attributes, beliefs, or opinions. In online social networks, this leads to preferential attachment between like-minded users, structurally reinforcing local agreement and limiting exposure to diverse viewpoints. As a result, homophily serves as a foundational mechanism behind the formation of \emph{echo chambers}, which in turn intensify \emph{polarization}, the divergence of opinions across distinct user groups.

To quantify the extent of homophilic connections, various metrics are used at different levels. One widely adopted metric is the \emph{External-Internal (EI) index}, defined as:
\[
EI = \frac{E - I}{E + I} \,,
\]
where $E$ is the number of edges linking a node or group to others with different attributes (external connections), and $I$ is the number of edges linking to those with similar attributes (internal connections). The EI-index ranges from $-1$ (perfect homophily) to $+1$ (perfect heterophily), with $0$ indicating an equal number of internal and external connections. A low or negative EI-index suggests that users are predominantly connected to similar peers, thus reinforcing ideological homogeneity and isolating them from diverse perspectives~\citep{bruns2017echo,kaiser2020birds}.

As a complementary homophily signal, \textit{assortativity coefficient} measures normalizely for a node attribute $x_i$ (e.g., ideology or stance), defined as \citep{Newman2018Networks}:

\begin{equation}
    r_{\mathrm{attr}}=\frac{\sum_{i j}(A_{i j}-k_ik_j/2m)x_ix_j}{\sum_{i j}(k_i\delta_{i j}-k_ik_j/2m)x_ix_j} \,.
\label{eq:modularity}
\end{equation}

Here $A_{ij}$ is the (possibly weighted) edge between $u_i$ and $u_j$, $k_i$ is the degree (or strength) of $u_i$, $m=\tfrac12\sum_{ij} A_{ij}$ is the total number of edges, $\delta_{ij}$ is the Kronecker delta, and $x_i$ encodes the attribute (e.g., $\pm 1$ label or a standardized scalar). The numerator captures attribute covariance across edges beyond the configuration-model baseline $k_i k_j/2m$; the denominator normalizes by the maximum variance under that baseline, yielding a Pearson-type correlation bounded in $[-1,1]$. Values $r_{\mathrm{attr}}>0$ indicate assortative (like-with-like) ties, $r_{\mathrm{attr}}<0$ indicate disassortative (cross-cutting) ties, and $r_{\mathrm{attr}}\approx 0$ is consistent with random mixing. For directed or weighted graphs, we use $A_{ij}$ as weights and may compute separate in- and out-versions by replacing $k_i$ with the in- or out-strengths to match the exposure direction.

However, while homophily captures pairwise similarity, it does not explicitly identify emergent clusters or communities in the network. Therefore, it is often complemented by community detection algorithms, such as Louvain and Infomap, which help uncover the structural boundaries that amplify homophilic tendencies and contribute to the formation of echo chambers.

\subsubsection{Controversy-based segregation}
While structural and interaction-based metrics shed light on community cohesion and node influence, they do not directly capture the extent to which a topic divides a network into opposing sides. Controversy-based segregation specifically quantifies the degree of polarization or fragmentation in a discussion, particularly in contexts such as political discourse or social issues. Such polarization reflects high ideological isolation, in which users from different viewpoints rarely interact or engage with one another, reinforcing echo chambers and deepening filter bubbles.

\textit{Random Walk Controversy (RWC)}
\citep{garimella2018quantifying} score is used to measure polarization in a two-partition network, where the graph is divided into two opposing communities, $X$ and $Y$, such that $X \cup Y = U$. The RWC is defined as: 
\begin{equation}
    \text{RWC} = P_{xx}P_{yy} - P_{xy}P_{yx}\,,
    \label{eq:rwc}
\end{equation}
where $P_{ab}$ denotes the probability that a random walk starting in community $a \in {X,Y}$ ends in community $b \in {X,Y}$.

High RWC (and variants) implies that random-walk exposure remains confined to like-minded communities, a direct operationalization of ideological isolation through constrained cross-ideological visibility.

Several extensions of the RWC have been proposed to refine the understanding of polarized interactions \citep{villa2021echo,munoz2024quantifying}:

\textit{Authoritative-RWC (A-RWC)} Focuses on walks between authoritative, emphasizing the separation of influential discourse leaders.
\[
  \text{A-RWC} = P^{(A)}_{xx} P^{(A)}_{yy} - P^{(A)}_{xy} P^{(A)}_{yx} \,,
  \label{eq:arwc}
\]
where \( P^{(A)}_{ab} \) denotes the probability that an authoritative node randomly walks starting from \(a\) and ends \(b\). Both are authoritative nodes in the community \( {X, Y} \).

\textit{Displacement-RWC (D-RWC)} Incorporates how information flows are displaced due to centrality-weighted movements across partitions, revealing more subtle dynamics of content segregation.

\[
\mathrm{DRWC}
= \frac{1}{|N|}
\sum_{v \in N}
\left( 1 - \frac{n(v)_{\text{cc}}}{l_{\text{rw}}} \right) \,,
\label{eq:drwc}
\]
where \( N \) is a sampled set of randomly selected vertices from the network (e.g., 60\% of each community), \( l_{\text{rw}} \) is the fixed length of the random walk, and \( n(v)_{\text{cc}} \) is the number of community changes encountered in the walk starting from node \( v \).

Higher D-RWC values indicate fewer inter-community transitions and stronger polarization.

\textit{Boundary Connectivity (BC)} \citep{villa2021echo} quantifies how boundary nodes split their ties between in-group and out-group neighbors. Let $B$ be the set of boundary nodes and $I$ the set of internal (same-side) nodes.
For a boundary node $u$, let $d_i(u)$ be the number of edges from $u$ to $I$, and $d_b(u)$ the number of edges from $u$ to $B$.
Then
\[
\mathrm{BC}
= \frac{1}{|B|}
\sum_{u \in B}
\left(
\frac{d_i(u)}{d_i(u)+d_b(u)} - 0.5
\right) \,,
\label{eq:bc}
\]
which lies in $[-0.5,\,0.5]$.
Values $\mathrm{BC}<0$ indicate little or no polarization; values $\mathrm{BC}>0$ mean boundary nodes, on average, preferentially connect inward (to internal same-side nodes) rather than across groups, signaling that controversy is likely present.

\textit{Betweenness centrality} \citep{mantzaris2014uncovering} captures a node’s role as a bridge between different parts of the network:
\[
C_B(u_i) = \sum_{s \neq i \neq t} \frac{\sigma_{st}(i)}{\sigma_{st}} \,,
\]
where $\sigma_{st}$ is the number of geodesic(shortest) paths from  $u_s$ to node $u_t$, and $\sigma_{st}(i)$ is the number of those paths passing through $u_i$. Nodes with high betweenness centrality can act as gatekeepers or facilitators of cross-group communication. When such bridging nodes are scarce, structurally peripheral, or bypassed in information flows, structural ideological isolation intensifies.

In these views, isolation arises when edges concentrate within like-minded subsets, boundary nodes preferentially attach inward, and random-walk mass becomes trapped within communities rather than diffusing across them. These signals quantify how the graph’s wiring constrains cross-ideological visibility, even when the network is globally connected.

\subsection{Content-based Metrics}
\label{subsec:content-metrics}

\subsubsection{Polarity and stance scores}
In the absence of explicit group labels (e.g., political affiliation or community membership), user ideology can be inferred from the content they post, share, or endorse. \emph{Polarity scores} and \emph{stance detection} offer effective means to estimate opinion leaning based on linguistic and emotional cues in user-generated content. These methods rely on sentiment analysis, political lexicons, or supervised classification models to assign a score or class label that reflects the user’s opinion orientation \citep{falkenberg2024patterns}.

Polarity scores ranging from “far-left” to “far-right” can be inferred from language features and retweet behavior in social media \citep{jiang2021social}. Let \( p_i \in [-1, 1] \) denote the polarity score of user \( u_i \), where \( -1 \) indicates one extreme (e.g., left-leaning or anti-stance) and \( +1 \) the opposite extreme (e.g., right-leaning or pro-stance). The polarity score can be computed using various strategies:
\begin{enumerate}
    \item Sentiment analysis of shared articles or posts.
    \item Aggregated stance of liked or re-posted content.
    \item Classifiers trained on labeled political datasets.
\end{enumerate}

We treat the scalar polarity as a one-dimensional belief, \(p_i \equiv b_i\).
Then the opinion distance between users \(u_i\) and \(u_j\) follows Eq.~(1) in Sec.~3:
\[
\Delta_{ij} = \left\lVert b_i - b_j \right\rVert = \left| p_i - p_j \right| \,,
\]
where a larger \( \Delta_{ij} \) indicates stronger divergence in stance. Aggregating these distances across a community or between communities can offer a quantitative view of ideological segregation.

In this context, content-based polarity and stance scores serve as proxies for user ideology, enabling researchers to detect echo chambers and filter bubbles through the lens of expressed opinions, even in the absence of network labels or explicit community structures.

\subsubsection{Entropy-based Diversity and Exposure}

This subsection quantifies informational breadth from two complementary angles: (i) \emph{topic/semantic diversity}, which characterizes the range of content that exists to be seen\citep{helberger2018exposure}, and (ii) \emph{exposure entropy}, which captures the distribution of what users actually see in their feeds\citep{hu2022ai}.

\textit{Topic Entropy (LDA-based):}
Let $\mathcal{T}=\{\tau_1,\ldots,\tau_K\}$ be the topic set, and let $\theta$ denote a topic-mixture distribution over $\mathcal{T}$ (e.g., from LDA). We define
\begin{equation}
H_{\text{topic}}
\;=\;
-\sum_{k=1}^{K} \theta(\tau_k)\,\log \theta(\tau_k) \,.
\tag{14}
\end{equation}
Lower $H_{\text{topic}}$ indicates a more concentrated (less diverse) thematic profile. This measure reflects the semantic opportunity set available in the corpus or in a user's candidate pool.

\textit{Semantic Entropy (BERT-based):} For a set of semantic vectors \( \{\mathbf{e}_1, \dots, \mathbf{e}_n\} \), diversity can be captured by computing pairwise cosine distances $d_{cos}$ and measuring their spread:
  \[
    H_{\text{semantic}} = \text{Entropy}\left( d_{cos}(\mathbf{e}_i, \mathbf{e}_j) \right)\,,
    \label{eq:semantic_entropy}
  \]
  where the entropy is computed over the distribution of pairwise distances.
  

\textit{Semantic Spread:} Defined as the average pairwise distance in embedding space:
  \[
    \mathrm{Spread} = \frac{2}{n(n-1)} \sum_{i<j} d_{cos}(\mathbf{e}_i, \mathbf{e}_j) \,,
    \label{eq:semantic_spread}
  \]
where a lower spread suggests semantically redundant content consumption.

 Consider content grouped into $K$ topics $\mathcal{T}=\{\tau_1,\ldots,\tau_K\}$ with exposure probabilities $p(\tau_k)$. The entropy of exposure is
\begin{equation}
H \;=\; -\sum_{k=1}^{K} p(\tau_k)\,\log p(\tau_k) \,.
\tag{15}
\end{equation}
At the user level $u_i$, let $p_i(\tau_k)$ denote the probability that $u_i$ is exposed to (or consumes) topic $\tau_k$; then
\[
H_{u_i} \;=\; -\sum_{k=1}^{K} p_i(\tau_k)\,\log p_i(\tau_k)\,.
\]
Lower $H_{u_i}$ signifies concentrated exposure, whereas higher values indicate a more even distribution across topics.

While $H_{\text{topic}}$ captures the semantic supply of potentially visible content, $H_{u_j}$ captures realized exposure. Ideological isolation is evidenced when both entropies are low, or when $H_{u_j}$ is markedly smaller than $H_{\text{topic}}$, implying that ranking mechanisms and/or user choices narrow attention to like-minded content despite broader semantic availability.

\subsubsection{Purity of Beliefs}
This metric evaluates the ideological homogeneity within a user community by assessing the proportion of users whose inferred ideology aligns with the community’s majority ideology~\citep{minici2022cascade}.

\begin{equation}
  \mathrm{Purity} = \frac{1}{|\mathcal{C}|} \sum_{u \in \mathcal{C}} \mathbb{I}[\hat{y}_u = y_{\text{majority}}] \,,
  \label{eq:purity}
\end{equation}
where \( \mathcal{C} \) is the set of users in a community (or echo chamber); \( \hat{y}_u \) is the inferred ideology label of user \( u \); \( y_{\text{majority}} \) is the dominant ideology label in \( \mathcal{C} \); \( \mathbb{I}[\cdot] \) is the indicator function that returns 1 if the condition is true, and 0 otherwise.

A higher purity score indicates greater ideological alignment among members, signaling semantic and ideological homogeneity within the group.

\subsubsection{Exposure bias and cross-cutting ratio}
A defining feature of echo chambers and filter bubbles is \emph{selective exposure}, in which users tend to encounter content that aligns with their preexisting beliefs. In contrast, opposing viewpoints are filtered out by algorithms or social behavior. To quantify this bias in content exposure, researchers measure the proportion of ideologically opposing (cross-cutting) content that a user encounters within their information stream, such as a social media feed or content recommendation list.

One of the most direct measures of this phenomenon is the \emph{exposure bias} \citep{bakshy2015exposure}, defined as:

\[
\text{EB}_i = 1 - \frac{C_i}{T_i} \,.
\]
For user \( u_i \), \( C_u \) be the number of ideologically cross-cutting posts encountered by user \( u \), \( T_u \) be the total number of posts seen by user \( u \).

An exposure bias value near 1 indicates highly selective exposure, suggesting the user is insulated from dissenting viewpoints, while values closer to 0 reflect balanced ideological exposure.

Conversely, the \emph{cross-cutting exposure ratio} can be defined simply as:
\[
\text{Cross-Cutting Ratio} = \frac{C_i}{T_i} \,,
\]
which serves as an inverse indicator of ideological isolation.

Low cross-cutting ratios or high exposure bias values are commonly interpreted as evidence of \emph{echo chamber} or \emph{tunnel vision} effects, where users are primarily exposed to information that reinforces their ideological views. This metric has been employed in influential studies, such as \citep{bakshy2015exposure}, which analyzed Facebook user behavior and found that both algorithmic filtering and user choice contributed to reduced cross-cutting exposure.

\subsection{User behavior Metrics}
\label{subsec:behavior-metrics}

\subsubsection{Interaction Symmetry and Diversity}

User behavior is a key driver of ideological isolation, particularly through interaction patterns that exhibit selective exposure \citep{srinath2025behavioral}. In digital environments, users often gravitate toward content that aligns with their beliefs, while avoiding opposing viewpoints, thereby reinforcing echo chambers and deepening confirmation bias. \textit{Interaction symmetry} aims to capture these behavioral tendencies by analyzing how users engage with the ideological diversity of the content presented to them \citep{bakshy2015exposure, ccetinkaya2025cross}.

The \textit{Engagement Symmetry} score for user \( u_i \), denoted as \( \text{ES}_i \), is defined as:

\[
\text{ES}_i = \frac{Int_i^{same} - Int_i^{opp}}{Int_i^{tot}} \,, Int_i^{tot}=Int_i^{same} + Int_i^{opp}
\]

\noindent
Here \(Int_i^{same} \) and \(Int_i^{opp} \)denote the (weighted) counts of \(u_i \)'s interactions with ideological aligned an opposing content respectively. A higher\( \text{ES}_i \) indicates stronger engagement with ideologically aligned content (echo chamber tendency), a lower \( \text{ES}_i \) indicates preference for opposing views, while values near zero reflect balanced engagement across ideological lines.

Symmetry reflects which side a user prefers, and entropy reflects how broadly they engage across stances or categories. 

To formalize this, entropy can be calculated over a user’s interaction distribution across content categories or ideological stances \citep{pecile2024decoding}. Given user $u_i$’s probability distribution of interactions across a set $\mathcal{C}$ of categories or topics at time $t$, denoted by $\{\,p^{(u_i)}_{c}(t)\,\}_{c \in \mathcal{C}}$, the The \textit{Interaction Entropy (IE)} is computed as:
\[
\mathrm{IE}_{u_i}(t) \;=\; - \sum_{c \in \mathcal{C}} p^{(u_i)}_{c}(t)\,\log p^{(u_i)}_{c}(t) \,.
\]
A high $\mathrm{IE}_{u_i}(t)$ indicates a user engages with a wide range of content, signaling openness and exposure diversity. In contrast, a low $\mathrm{IE}_{u_i}(t)$ implies behavioral concentration, suggesting that the user may be entrenched in a narrow ideological niche.

\subsubsection{Temporal Reinforcement}

Temporal reinforcement refers to the phenomenon where recommender systems gradually deliver increasingly similar content to users over time, thereby reinforcing users’ prior preferences and narrowing their information exposure. This phenomenon can be quantified by observing changes in content diversity across recommendation sequences. 

Ge et al. demonstrate this effect by analyzing the Euclidean distance between the embeddings of consecutively recommended items \citep{ge2020understanding}. In a high-dimensional semantic space (e.g., BERT or collaborative filtering embeddings), each item is represented as a vector. The Euclidean distance between successive items \( \mathbf{x}_t \) and \( \mathbf{x}_{t+1} \) is defined as:

\[
d_t = \|\mathbf{x}_{t+1} - \mathbf{x}_t\|_2 \,.
\]

This temporal narrowing contributes directly to the emergence of echo chambers and filter bubbles, and can culminate in tunnel vision—prolonged selective exposure that concentrates attention on a single aspect of an event, further reducing viewpoint diversity and amplifying ideological isolation. Through algorithmic filtering and behavioral feedback loops, personalization increasingly homogenizes future recommendations, increasing temporal stability of exposure and reinforcing confirmation bias and polarization. 

To quantify this stability, researchers track changes in exposure distributions (e.g., topics, sentiments, ideologies) over time. A notable decline in diversity suggests an increase in informational confinement and the strengthening of echo chambers. One common approach to measure this temporal change is the use of \emph{Kullback-Leibler (KL) divergence}, which quantifies how much one probability distribution diverges from another\citep{bishop2006pattern}. 

Given a user $u_i$'s exposure distributions $P_i(t)$ and $P_i(t+1)$ at consecutive timepoints $t$ and $t+1$, the KL-divergence is defined as:
\[
D_{\text{KL}}(P_i(t) \| P_i(t+1)) = \sum_{k=1}^K P_i(t,\tau_k) \log \frac{P_i(t,\tau_k)}{P_i(t+1,\tau_k)} \,,
\]
where $\tau_k$ denotes topic k, and $P_i(t)$ is the probability that $u_i$ is exposed to $\tau_k$ at time $t$. A high $D_{\text{KL}}$ indicates a significant shift in the user’s content exposure between timepoints, which may signal exploration or a change in user behavior. In contrast, a persistently low $D_{\text{KL}}$ indicates temporal stability, often reflects algorithmic reinforcement, and increases the homogeneity of content consumption.

Additionally, to track the overall diversity of exposure over time, we use the entropy definition in Eq.~(15). We track temporal change via
\[
\Delta H_{u_i}(t)=H_{u_i}(t{+}1)-H_{u_i}(t), 
\qquad 
\widehat{H}_{u_i}(t)=\frac{H_{u_i}(t)}{\log K} \,.
\]

A persistently decreasing $\widehat{H}_{u_i}(t)$ together with low but stable 
$D_{\mathrm{KL}}$ indicate temporal reinforcement and narrowing exposure, consistent with entropy-based analysis of echo chambers and selective exposure
\citep{cinelli2020echo}.

\subsubsection{Like-based Polarization}

Like-based polarization is a user-level metric that quantifies ideological alignment based on observable engagement behaviors, specifically the number of "likes" on social media platforms \citep{brugnoli2019recursive}. In many online environments, users are exposed to content representing opposing narratives. For example, scientific information versus conspiracy theories, or left-leaning versus right-leaning political discourse.

Brugnoli et al. \citep{brugnoli2019recursive} propose a simple yet effective polarization score to capture this behavior:

\begin{equation}
  \rho(u) \;=\; \frac{x - y}{x + y} \,,
  \label{eq:brugnoli}
\end{equation}
where  \( x \) denotes the number of likes a user \( u \) gives to content aligned with one narrative (e.g., conspiracy); \( y \) refers to the number of likes given to content from the opposing narrative (e.g., science); The resulting score \( \rho(u) \in [-1, 1] \) reflects the user’s ideological leaning: A higher \( \rho(u) \) indicates stronger alignment with one ideological side (e.g., conspiracy), 
a lower \( \rho(u) \) indicates alignment with the opposing side (e.g., science), 
and values near zero represent balanced engagement across both narratives.

A bimodal distribution of \( \rho(u) \) values across the population, with peaks near \(-1\) and \(1\),  indicates the presence of two polarized user groups, with few users engaging meaningfully with both sides. This behavioral pattern reflects the fragmentation of public discourse into ideologically homogeneous communities, a hallmark of echo chambers.

Moreover, this metric directly connects to filter bubbles: as platforms personalize content based on prior likes, users become increasingly surrounded by narratives that reinforce their initial preferences. The result is a feedback loop that magnifies polarization at both the individual and community levels.

\subsubsection{Bimodality and balance}
Bimodality and balance are statistical metrics used to capture the degree and structure of opinion divergence in a population \citep{pfister2013good}. In online discourse, bimodality refers to a distribution of user opinions with two distinct peaks, indicating the presence of two dominant ideological camps. This suggests that users are not uniformly distributed across the opinion spectrum but instead cluster around opposing poles. Such separation signifies strong \emph{behavioral polarization}, where engagement patterns and content preferences reflect entrenched ideological positions.

Bimodality is often accompanied by asymmetric group sizes, with one ideological camp significantly outnumbering the other. This asymmetry is captured through the notion of \emph{balance}, which quantifies the relative size of opposing opinion groups. When combined, bimodality and balance offer a nuanced view of polarization, revealing not only the separation between camps but also the structural marginalization that may lead to ideological isolation or filter bubble effects.

To quantify bimodality in opinion distributions, the \emph{Bimodality Coefficient (BMC)} \citep{pfister2013good,bessi2016users} is used:
\[
\text{BMC}(t) = \frac{\gamma_1(t)^2 + 1}{\gamma_2(t) + \frac{3(n(t)-1)^2}{(n(t)-2)(n(t)-3)}} \,,
\]
where $\gamma_1(t)$ is the skewness, $\gamma_2(t)$ is the kurtosis, and $n(t)$ is the number of users at time $t$. A value of $\text{BMC}(t) > 0.55$ typically indicates a bimodal or skewed distribution \citep{freeman2013assessing}. In online settings, this often corresponds to two ideologically homogeneous camps with few users expressing moderate views. As such, a high bimodality coefficient is interpreted as a strong signal of opinion polarization and the emergence of echo chambers.

Complementing the separation captured by the bimodality coefficient $\text{BMC}(t)$, the \emph{Balance Ratio (BR)} assesses the symmetry of group sizes by taking the ratio of the smaller to the larger of the two dominant clusters\citep{huang2021deep}:
\[
\mathrm{BR}(t)=\frac{\min\{\,|\mathcal{G}_1(t)|,\;|\mathcal{G}_2(t)|\,\}}{\max\{\,|\mathcal{G}_1(t)|,\;|\mathcal{G}_2(t)|\,\}} \,,
\]
where $\mathcal{G}_1(t)$ and $\mathcal{G}_2(t)$ are the two dominant ideological clusters detected at time $t$.
The balance ratio lies in $(0,1]$: values near $1$ indicate equally sized camps (symmetric polarization), while lower values indicate a dominant majority versus a much smaller minority. A low balance ratio heightens ideological isolation for the minority group, as fewer like-minded peers and items constrain within-camp exposure and support, thereby increasing minority social/structural isolation.

\subsubsection{Cascade shape metrics}

Information cascades, chains of resharing or reposting, leave structural footprints of diffusion that reveal whether engagement is broad or confined to like-minded communities \citep{cinelli2021echo}. Cascade shape metrics quantify these dynamics through depth, breadth, structural virality, and density, offering a behavioral–structural lens on ideological isolation and selective propagation. Deep, cross-community cascades suggest broad exposure, whereas shallow, wide cascades confined to tightly connected clusters indicate echo-chamber reinforcement. We next introduce four key metrics.

\textit{Cascade Depth ($D$) \citep{leskovec2007cost}} refers to the longest path from the original sharer to any node in the cascade. It captures how far content travels hierarchically:
\[
D = \max_{u \in U} \text{depth}(u) \,,
\]
where $U$ is the set of nodes in the cascade tree.

\textit{Cascade Breadth ($B$) \citep{leskovec2007cost}} indicates the maximum number of nodes at any single level of the cascade:
\[
B = \max_{l} |U_l| \,,
\]
where $U_l$ is the set of nodes at level $l$ of the cascade.

\textit{Structural Virality ($SV$) \citep{goel2016structural}} refers to the average pairwise distance between all nodes in the cascade. This metric balances both depth and breadth:
\[
SV = \frac{1}{n(n-1)} \sum_{i \neq j} d(i,j) \,,
\]
where $n$ is the number of nodes in the cascade and $d(i,j)$ is the shortest path between nodes $u_i$ and $u_j$.

\textit{Cascade Density \citep{Newman2018Networks}} measures the ratio of actual to possible edges in the cascade graph. A high density indicates intense local resharing:
\[
\text{Density} = \frac{2| E|}{|U|(|U| - 1)} \,,
\]
where $ E$ and $U$ are the edge and node sets, respectively.

Low structural virality, coupled with high density and shallow depth, is typical of cascades confined within echo chambers, where content is predominantly reshared among a homogeneous group. In contrast, cascades that reach high depth and exhibit low density are more likely to cross ideological or community boundaries, reflecting greater diversity in content exposure.

\section{Computational Mitigation Strategies for Ideological Isolation}
\label{sec:mitigation_strategies}

This section focuses on computational approaches designed to mitigate ideological isolation in online social networks. As illustrated in Figure 3, the mitigation strategies discussed here correspond directly to the metric families and isolation types introduced in Section 4, thereby providing an operational link between diagnosis and intervention. We organize the discussion into three complementary perspectives. User modeling and simulation approaches replicate belief dynamics and user interactions to evaluate the effectiveness of interventions. Network-topological approaches aim to reshape connectivity by adding, removing, or reweighting edges to enhance cross-cutting exposure. Content-based control mechanisms intervene within recommendation and ranking pipelines through pre-, in-, and post-processing to promote informational diversity while maintaining relevance. Together, these strategies form a computational toolkit for reducing the self-reinforcing mechanisms that sustain ideological isolation.

\subsection{User Modeling and Simulation}

To develop strategies for mitigating ideological isolation, it is essential to understand and model users' behaviors, i.e., how users respond to the information. Such modeling and simulation not only capture the dynamics of user interaction and belief evolution but also provide feedback for evaluating the effectiveness of the proposed strategies \citep{de2022modelling}. In this subsection, we survey several popular approaches for modeling and simulating users' behaviors. 

\subsubsection{Bayesian Modeling}

To model how users perceive and disseminate information in social networks, Bayesian frameworks have been widely adopted to capture the belief-updating process driven by interactions with neighbors \citep{masterton2013argumentation,jimenez2015model}. 

Generally, each user \( u \) maintains a prior belief distribution over potential message intents or themes, denoted by \( \{m_i\} \). This prior, \( P(m_i) \), reflects the user's historical preferences or typical communicative behavior. At each timestep, the user observes recent messages from their neighbors, denoted by \( \mathbf{M}_{\varGamma(u)} \), where \( \varGamma(u) \) is the set of user \( u \)'s neighbors. These neighbor messages provide evidence that may influence users' beliefs and guide their response generation.
Using Bayes’ theorem, the user updates their belief distribution over possible messages as:
\[
P(m_i \mid \mathbf{M}_{\varGamma(u)}) \propto P(m_i) \cdot P(\mathbf{M}_{\varGamma(u)} \mid m_i) \,,
\]
where \( P(m_i) \) is the prior probability of producing message \( m_i \), and \( P(\mathbf{M}_{\varGamma(u)} \mid m_i) \) denotes the likelihood of observing the neighbor messages \( \mathbf{M}_{\varGamma(u)} \) given intent \( m_i \). This likelihood can be estimated based on semantic similarity, topical relevance, or discourse coherence between \( m_i \) and the neighbor messages.

This Bayesian mechanism enables users to adapt their views based on local interactions and to propagate information aligned with their updated beliefs, thereby simulating a dynamic social information flow.

\subsubsection{Agent-Based Modeling and Simulation}

Agent-Based Models (ABMs) are widely used to study the emergence of ideological isolation by explicitly representing individual traits and behaviors. A critical aspect of these models is understanding how agents influence, and are influenced by, their neighbors \citep{bastarianto2023agent,Zhang2015HIPRank}. Prior research highlights personal preference as a key factor, alongside the influence of surrounding agents \citep{li2019multi,wu2019adaptive}. To capture both intrinsic predispositions and social interaction effects, two mechanisms, \emph{Prior Commitment Level} (PCL) and \emph{Comprehensive Social Influence} (CSI), are incorporated into ABM frameworks \citep{li2019multi,Zhang2015HIPRank}.

Mathematically, each agent $u_i$ is part of a social network $G=(V,E)$ with a set of neighbors $\varGamma(u_i)$. The PCL $pcl_{jx}$ measures agent $u_j$'s predisposition toward item $i_x$ based on past ratings, normalized as:
\begin{equation}
pcl_{jx} =
\begin{cases}
\frac{r_{jx} - \min(R_j)}{\max(R_j) - \min(R_j)}, & \max(R_j) \neq \min(R_j) \\
0.5, & \text{otherwise}
\end{cases}
\end{equation}
where $R_j$ is the set of ratings given by $u_j$. Higher $pcl_{jx}$ implies stronger favorability, while lower values indicate the opposite. The CSI reflects the cumulative persuasive strength from neighbors $\varGamma(u_j)$ who hold a given opinion:
\begin{equation}
csi_{jx} = 1 - \prod\nolimits_{v_m \in \varGamma(u_j)} (1 - ipp_{mj}) \,,
\end{equation}
where $ipp_{mj}$ is the interpersonal persuasion power from $v_m$ to $u_j$. The probability of $u_j$ adopting a new opinion is a weighted combination of personal predisposition (PCL) and social influence (CSI), modulated by a trade-off parameter $\lambda_j \in [0,1]$:
\begin{equation}
p_{jx} = \lambda_j \cdot pcl_{jx} + (1 - \lambda_j) \cdot \frac{csi_{jx}}{\sum\nolimits csi_{jx'}} \,.
\end{equation}
This formulation allows ABMs to capture how individual tendencies and peer influence jointly drive opinion dynamics.

Another approach to designing agents in ABM is to define them through a set of explicit rules that govern how they perceive, process, and act upon information \citep{li2021social,wang2023maximizing}. At each time step $t$, it receives \emph{influence messages} $msg_p(u_j \rightarrow u_i,t)$ from neighbors $u_j \in \varGamma(u_i)$, where each message carries topical relevance and an opinion indicator.

Formally, a message $msg_p$ can be represented as:
\begin{equation}
S_{msg_p} = \{ m_p(\tau_k) \mid \tau_k \in T \}, \quad m_p(\tau_k) \in [0,1] \,,
\end{equation}
where $m_p(\tau_k)$ is the degree of association with topic $\tau_k$.  
The agent retains a posting history $PM^{(u_i)}$ and a received message set $I^{(u_i)}$, which together shape its current context $C_L(u_i,t)$. Decision-making is then modeled as:

\begin{equation}
\text{NextMessage}(u_i,t) = f\big(I^{(u_i)}, PM^{(u_i)}, \varGamma(u_i)\big) \,,
\end{equation}
where $f(\cdot)$ encodes behavioral rules (e.g., Perceive recent messages from neighbors, Incorporate own past expressed opinions, Select and deliver the next influence message).

By leveraging ABM, the network's evolution is driven by individual agents' behaviors \citep{li2016agentbasednetworks}. Over time, this evolution can reveal ideological patterns, which can be evaluated using established metrics of ideology. 

\subsubsection{LLM-Driven Simulation}

An emerging direction in ABMs is the integration of large language models (LLMs) to simulate realistic, context-aware user behaviors~\citep{hu2024llm,zhang2025llm}. In this paradigm, each agent $u_i$ operates within a social network $G$ and maintains repositories of received messages $R_i^{in}$ and posted messages $R_i^{out}$. At each timestep $t$, an agent decides whether to broadcast, adopt, or evolve a message based on probabilistic thresholds. When triggered, the agent uses an LLM as a function to generate a response $m_x$ conditioned on both the message content $c_x$ and the agent’s profile $u_i$: 
\[
m_x = \mathrm{LLM}(c_x, u_i) \,.
\]

This mechanism enables the simulation of nuanced discourse generation, where responses are shaped by both the social context (e.g., influence from neighbors $\varGamma(u_i)$) and the agent’s prior outputs. By iteratively applying
\[
R_i^{out}(t+1) = R_i^{out}(t) \cup \{ m_x \}, \quad R_j^{in}(t+1) = R_j^{in}(t) \cup \{ m_x \}
\]
for targeted neighbors $u_j$, the model captures message propagation, evolution, and conversational focus shifts over time. Compared with traditional probability-based ABMs, this approach enables agents to produce language-rich, semantically coherent interactions, thereby allowing the study of how influence diffusion shapes collective discourse and topic dominance in online environments. On the other hand, because LLM behavior is more uncertain and its use requires greater time and computational resources, applying LLM-based Monte Carlo simulation to large-scale networks is challenging. However, if the model can tolerate variability in individual behaviors, LLM-driven simulation remains a suitable option.

\subsubsection{Reinforcement Learning}

A complementary way to model users is to treat each user agent as a reinforcement-learning (RL) decision maker embedded in a social network, as RL naturally captures the sequential, feedback-driven nature of online interactions, allowing agents to adapt their strategies over time to maximize long-term engagement or influence \citep{sert2020segregation,song2025online,wu2023gac}. Mathematically, at each time step $t$, agent $u_i$ receives an observation:
\[
o_t^{(i)} = g\!\big(M_t^{(i)},\, x_t^{(i)},\, c_t\big) \,,
\]
where $M_t^{(i)}$ is the set of messages from neighbors $\Gamma(u_i)$ at timestep $t$, $x_t^{(i)}$ encodes the agent’s internal state (e.g., preferences/history), and $c_t$ denotes contextual features (topic, time, forum). The agent selects an action from a finite set: 
\[
a_t^{(i)} \in \mathcal{A}=\{\text{ignore},\ \text{reshare},\ \text{reply},\ \text{click},\ \text{compose}, ...\} \,,
\]
according to a policy $\pi_\theta(a\mid o)$ (e.g., Deep Q-Learning (DQN) or actor–critic).

The immediate utility of an action is captured by a reward function that balances engagement, preference alignment, and effort/risk:
\[
r_t^{(i)} \;=\; \alpha\,E\!\big(m_t^{(i)}\big)\;+\;\beta\,S\!\big(m_t^{(i)},P^{(i)}\big)\;-\;\eta\,C\!\big(a_t^{(i)}\big) \,,
\]
where $m_t^{(i)}$ is the produced/forwarded message, $E(\cdot)$ measures observable engagement (e.g., click, reply), $S(\cdot, P^{(i)})$ measures semantic alignment with the agent’s preference profile $P^{(i)}$, and $C(\cdot)$ penalizes effort, delay, or moderation risk. To reflect social spillovers, the reward can include a neighborhood term:
\[
\bar r_t^{(i)} \;=\; r_t^{(i)} \;+\; \lambda \sum_{j\in\Gamma(u_i)} w_{ij}\,\Delta \Psi_t^{(j)} \,,
\]
where $\Delta \Psi_t^{(j)}$ is a change in neighbor $u_j$’s engagement or stance and $w_{ij}$ is an influence weight. The objective is to maximize the discounted return
\[
J(\theta)\;=\;\mathbb{E}_{\pi_\theta}\!\left[\sum_{t=0}^{T}\gamma^{t}\,\bar r_t^{(i)}\right] \,,
\]
where $\gamma \in [0,1]$ denotes the discount factor, and $J(\theta)$ is optimized either via value-based learning with the Bellman target:
\[
Q(o_t^{(i)},a_t^{(i)}) \;\leftarrow\; r_t^{(i)}+\gamma\,\max_{a'} Q(o_{t+1}^{(i)},a')
\]
or via policy gradient:
\[
\nabla_\theta J(\theta)\;=\;\mathbb{E}\!\left[\nabla_\theta \log \pi_\theta(a_t^{(i)}\!\mid o_t^{(i)})\,(G_t^{(i)}-b(o_t^{(i)}))\right] \,,
\]
with return $G_t^{(i)}$ and baseline $b(\cdot)$ for variance reduction. Message diffusion is modeled by updating neighbors’ inboxes stochastically, e.g.,
\[
R_{j}^{\text{in}}(t{+}1)\;=\;R_{j}^{\text{in}}(t)\,\cup\,\{m_t^{(i)}\}\quad\text{with probability }p_{ij} \,.
\]

Because observations are partial, $o_t^{(i)}$ is often encoded with a history aggregator (RNN/transformer) to approximate a belief state. This RL formalization yields agents that \emph{read} incoming messages, \emph{decide} among response options, and \emph{adapt} behavior over time to maximize long-term social utility, providing a principled alternative to purely rule- or probability-based ABMs.

\subsubsection{Graph Neural Networks}

Graph Neural Networks (GNNs) provide a powerful framework for modeling user behaviors in online social networks, where agents (users) are naturally represented as nodes and their connections as edges \citep{qiu2018deepinf,wu2018socialgcn}. In this setting, each agent $u_i$ receives messages from its neighbors $\Gamma(u_i)$, processes them according to its internal state, and decides whether and what message to deliver in the next step. GNNs capture both the structural information of the network and the evolving states of users, enabling the simulation or prediction of realistic message-propagation patterns.

Specifically, at time $t$, each user $u_i$ has a feature vector $\mathbf{h}_i^{(t)}$ representing its current state (e.g., preference profile, recent message embeddings). Incoming messages from neighbors can be encoded into message vectors $\mathbf{m}_{ij}^{(t)}$, aggregated as:
\begin{equation}
\mathbf{m}_i^{(t)} = \mathrm{AGGREGATE}\left(\left\{ \mathbf{m}_{ij}^{(t)} \mid u_j \in \Gamma(u_i) \right\}\right) \,,
\end{equation}
where $\mathrm{AGGREGATE}(\cdot)$ can be a mean, sum, or attention-weighted sum. The node update rule in a GNN layer combines the agent’s current state and aggregated neighbor messages:
\begin{equation}
\mathbf{h}_i^{(t+1)} = \sigma\!\left( W_1 \mathbf{h}_i^{(t)} + W_2 \mathbf{m}_i^{(t)} \right) \,,
\end{equation}
where $W_1$ and $W_2$ are learnable weight matrices and $\sigma(\cdot)$ is a non-linear activation function. In attention-based GNNs, $\mathbf{m}_i^{(t)}$ can be computed using:
\begin{align}
& \alpha_{ij} = \frac{\exp\!\left( \mathrm{LeakyReLU}(\mathbf{a}^\top [W\mathbf{h}_i^{(t)} \,\|\, W\mathbf{h}_j^{(t)}]) \right)}
{\sum_{k \in \Gamma(u_i)} \exp\!\left( \mathrm{LeakyReLU}(\mathbf{a}^\top [W\mathbf{h}_i^{(t)} \,\|\, W\mathbf{h}_k^{(t)}]) \right)} \,, \\
& \mathbf{m}_i^{(t)} = \sum_{j \in \Gamma(u_i)} \alpha_{ij} \, W \mathbf{h}_j^{(t)} \,,
\end{align}
where $\alpha_{ij}$ is the attention weight representing the relative influence of neighbor $u_j$ on $u_i$.

The decision to deliver a new message is then modeled as a classification or generation problem:
\begin{equation}
p(a_i^{(t)} \mid \mathbf{h}_i^{(t+1)}) = \mathrm{softmax}\!\left( W_o \mathbf{h}_i^{(t+1)} \right),
\end{equation}
where $a_i^{(t)}$ denotes the action (e.g., ignore, reshare, modify, or create a new message) and $W_o$ is a learnable parameter matrix. 

In predictive settings, the GNN is trained to maximize the likelihood of observed actions; in simulation settings, $a_i^{(t)}$ is sampled from $p(\cdot)$ to update the network state. By iteratively updating node states and simulating message exchanges, GNN-based models can reproduce or forecast complex message diffusion behaviors that depend on both network topology and user-level dynamics, outperforming purely probabilistic or rule-based models in capturing higher-order interaction effects \citep{fan2019graph,wu2020comprehensive}.

\subsection{Network-Topological Approaches}

From a network topology perspective, two primary strategies are commonly employed to mitigate ideological isolation and enhance diversity of information exposure:  
(1) \textit{Edge modifications}, and  
(2) \textit{Node modifications} (via influence spread).  

\subsubsection{Edge Modifications}

Edge modifications involve \emph{adding}, \emph{removing}, or \emph{reweighting} specific edges in the network to alter existing interaction patterns. These interventions are designed to weaken the structural mechanisms, such as community boundaries or echo chambers, that reinforce ideological isolation.

\paragraph{\textbf{Edge Addition}}
Adding cross-partition edges, links between nodes in opposing communities \(X\) and \(Y\), can weaken segregation by increasing cross-group transition probabilities (\(P_{xy}, P_{yx}\)) and decreasing within-group persistence (\(P_{xx}, P_{yy}\)). Using the RWC definition in Equation \ref{eq:rwc}, the edge-addition problem can be formulated as minimizing polarization after augmenting the graph with a small set of cross-cutting links~\citep{garimella2017reducing}:
\[
\mathrm{RWC}(G) \;=\; P_{xx}(G)\,P_{yy}(G)\;-\;P_{xy}(G)\,P_{yx}(G),
\]
\[
\min_{\Delta E \subseteq (X\times Y)\cup(Y\times X)} \;\mathrm{RWC}\!\left(G \oplus \Delta E\right)
\quad \text{s.t.} \quad |\Delta E|\le k,
\]
where \(G \oplus \Delta E\) denotes the graph after adding the directed edges \(\Delta E\). A practical strategy is greedy selection by marginal gain:
\[
\Delta\mathrm{RWC}(u,v)\;=\;\mathrm{RWC}(G)\;-\;\mathrm{RWC}\!\left(G \oplus \{(u,v)\}\right),
\]
choosing up to \(k\) candidate edges \((u,v)\in X\times Y \cup Y\times X\) that yield the largest decrease in RWC. Depending on the mitigation goal, the same objective can be extended to target influential nodes by replacing \(\mathrm{RWC}\) with \(\mathrm{A\!-\!RWC}\), or to discourage long runs within a single community by replacing it with \(\mathrm{D\!-\!RWC}\), thereby prioritizing edges that specifically reduce elite-level separation or increase inter-community transitions along random walks.

Building on this formulation, several heuristics and optimization methods have been proposed to identify effective cross-community links that minimize RWC or related polarization indices. Typical methods include: 
\subparagraph{High-degree bridging} \citep{garimella2017reducing} 
Given a partition of Graph $G$, i.e., \(V'=\{X \cup Y\}\) with \(\{X \cap Y\}=\varnothing\), define the candidate cross-community edges 
\[
 E_\times \;=\; \{(u,v)\in X\times Y \;:\; (u,v)\notin E\} \,.
\]
A simple and effective scoring rule is the degree-product:
\[
s_{\deg}(u,v) \;=\; k(u)\,k(v),
\]
where \(k(\cdot)\) is node degree. Select the top-\(k\) edges by \(s_{\deg}\). 
This favors bridges between structurally influential nodes, which (i) increases the cut size between \(X\) and \(Y\) by \(k\), (ii) reduces modular segregation, and (iii) raises cross-community random-walk flow. 

\subparagraph{Greedy selection for edge addition} Let \(F(G)\) be a polarization objective to \emph{minimize} (e.g., RWC in Equation \ref{eq:rwc} or modularity \(Q\) in Equation \ref{eq:modularity}). 
Define the marginal gain of adding a candidate edge \(e\) to the current graph \(G_t\) as
\[
\Delta_F(e\,|\,G_t) \;=\; F(G_t)\;-\;F(G_t + e) \,.
\]
The greedy rule (for a budget of \(k\) edges) is
\[
e_t^\star \;=\; \arg\max_{e\in E_\times} \,\Delta_F(e\,|\,G_{t-1}) \,, 
\]
\[
G_t \leftarrow G_{t-1}+e_t^\star,\quad t=1,\dots,k \,.
\]

\paragraph{\textbf{Edge Removal}} A line of work studies \emph{removing} a small number of \emph{intra-community} links to weaken echo chambers while preserving overall connectivity. Recall that opinion dynamics are often modeled with the Friedkin--Johnsen (FJ), which is introduced in Equation \ref{eq:FJ}. 

The task is commonly cast as selecting a small deletion set \(\Delta E\) (budget \(k\)) that minimizes a chosen \emph{echo-chamber index} \(\mathrm{EC}(G)\), subject to structural constraints:
\[
\min_{\Delta E \subseteq E_{\mathrm{intra}},\,|\Delta E|\le k}\; \mathrm{EC}\big(G\setminus \Delta E\big) \,,
\]
subject to $(G\setminus \Delta E)$ remains connected and inter-community edges are retained. 
This formulation summarizes existing approaches that target reinforcing links rather than bridges, aiming to attenuate echo-chamber effects without fragmenting the network. 

\subparagraph{Targeted removal of high-betweenness intra-community edges}
Let $b(e)$ be the \emph{edge betweenness} (number of shortest paths using $e$) \citep{girvan2002community}:
\[
b(e)\;=\;\sum_{s\neq t}\frac{\sigma_{st}(e)}{\sigma_{st}} \,,
\]
where $\sigma_{st}$ is the number of shortest $s{\to}t$ paths and $\sigma_{st}(e)$ those that traverse $e$.
Restrict candidates to $E_{\text{intra}}=\{(u,v)\in E:\,c_u=c_v\}$ and (under a connectivity constraint) remove the top-$k$ edges by $b(e)$.
The intuition is that deleting intra-community "highways" weakens insular clusters without destroying cross-community bridges. 

\subparagraph{Greedy conflict minimization} \citep{chen2018Quantifying} 
Let $R(G)$ denote a conflict/polarization \emph{risk} (e.g., the average-case risk under Friedkin--Johnsen dynamics).
At each step, remove the \emph{intra}-community edge with the most significant marginal risk drop:
\[
e^\star \;=\;\arg\max_{e\in E_{\text{intra}}}\Big(R(G)\;-\;R(G\setminus\{e\})\Big) \,.
\]
A concrete instantiation uses steady-state opinions $\mathbf{x}^\ast(G)$ from the FJ model and an intra-edge disagreement objective
\[
D_{\text{intra}}(G)\;=\;\sum_{(i,j)\in E_{\text{intra}}}\!\!\big(x_i^\ast-x_j^\ast\big)^2 \,,
\]
then greedily deletes edges maximizing $D_{\text{intra}}(G)-D_{\text{intra}}(G\setminus\{e\})$, subject to preserving connectivity. 
This is based on the intuition that removing links that most reduce within-group reinforcement. 

\subparagraph{Structural-bias (random-walk) reduction} \citep{haddadan2021repbublik} 
Let $V_{\mathrm{opp}(u)}$ be the opposite side of $u$. Define the \emph{polarized bubble radius} (expected hitting time to the opposite side)
\[
\mathrm{BR}(u)\;=\;\mathbb{E}\!\big[\tau_{u\to v}\,\big|\,v\in V_{\mathrm{opp}(u)}\big] \,,
\]
and the global structural bias $\Phi(G)=\sum_{u\in V}\mathrm{BR}(u)$.
Select intra-community deletions that maximize the reduction
\[
\Delta_\Phi(e)\;=\;\Phi(G)\;-\;\Phi\!\big(G\setminus\{e\}\big) \,,
\]
again with connectivity safeguards. 
This is based on the intuition that deleting edges that keep random walks “circulating” inside one side shortens stochastic access to opposing views.

\paragraph{\textbf{Edge Edition}} Edge editing adjusts the weights of existing links to influence the dynamics of opinion. This approach fine-tunes interpersonal influence strengths, simulating how recommended systems or moderation algorithms amplify or suppress interactions. 

Chitra and Musco augment the FJ opinion model by introducing a network administrator that adaptively reweights edges to minimize disagreement, and an algorithmic filtering mechanism that can induce filter bubbles \citep{chitra2020analyzing}. In round $r$, users first update to the FJ equilibrium on the current graph:
\[
z^{(r)} \;=\; \big(L^{(r-1)} + I\big)^{-1}s \,,
\]
where $L^{(r-1)}$ is the Laplacian, $s$ are innate opinions, and $z$ are expressed opinions. Disagreement is
\[
D_{G,z} \;=\; z^\top L z \,.
\]
Given $z^{(r)}$, the administrator then chooses a new graph (i.e., edge weights $W$) by
\[
G^{(r)} \;=\; \arg\min_{G\in \mathcal{S}} \; D_{G,\,z^{(r)}} \,,
\]
subject to realistic constraints $\|W-W^{(0)}\|_F \le \varepsilon \|W^{(0)}\|_F$ and row-sum (degree) preservation $\sum_j W_{ij}=\sum_j W^{(0)}_{ij}$ for all $i$. Alternating these steps can \emph{increase polarization dramatically} on real Reddit/Twitter graphs; a simple regularized variant (adding $\gamma\|W\|_F^2$) limits the filter-bubble effect while barely affecting disagreement.

\subsubsection{Node Modifications}
Node-level interventions to mitigate ideological isolation typically follow two approaches: (i) modifying existing node features, such as susceptibility, thresholds, or receptivity, to influence how opinions spread, often within an influence maximization framework; and (ii) introducing or designating special nodes, such as seeds or stubborn "mediator" agents, to inject cross-cutting information. The former is more prevalent in the literature, while the latter offers a complementary strategy for disrupting echo chambers. Both approaches are based on the social influence diffusion models. 

\paragraph{\textbf{Social Influence Diffusion}}
Social influence diffusion models describe how behaviors, opinions, or information propagate through networks. Two widely used frameworks are the \emph{Independent Cascade} (IC) and \emph{Linear Threshold} (LT) models \citep{kempe2003maximizing}.

In the \textit{IC model}, each directed edge $(u_j,u_i)$ has an activation probability $p_{ji} \in [0,1]$.  
Let $A_t$ be the set of active nodes at time $t$. The probability that an inactive node $u_i$ becomes active at $t+1$ is:
\[
P\big(u_i \ \text{activates at} \ t+1\big) 
= 1 - \prod_{u_j \in A_t \cap \varGamma(u_i)} \big(1 - p_{ji}\big) \,,
\]
where $\varGamma(u_i)$ is the set of in-neighbors of $u_i$.
Once active, a node remains active for all subsequent steps.

In the \textit{LT model}, each node $u_i$ has a threshold $\theta_u \in (0,1]$ and receives influence weights $w_{u_ju_i}$ from its neighbors, with $\sum_{u_j\in \varGamma(u_i)} w_{u_ju_i} \le 1$. At each step, an inactive node $u_i$ becomes active if the total influence from its active neighbors reaches or exceeds its threshold:
\[
\sum_{u_j \in \{\varGamma(u_i) \cap A\}} w_{u_iu_j} \ \ge \ \theta_v \,.
\]
As in the IC model, activation is irreversible.

Both models naturally capture the stochastic and progressive nature of influence and are submodular under common assumptions, enabling efficient seed selection through greedy algorithms with approximation guarantees \citep{kempe2003maximizing}. In the context of breaking echo chambers, IC and LT can be used to model the spread of interventions, such as pro-diversity content or belief adjustments, across communities.

\paragraph{\textbf{Susceptibility tuning via influence maximization}}
The key idea is to target highly insular communities with interventions that lower their susceptibility to within-group influence, thereby weakening echo chambers and promoting cross-cutting exposure \citep{panagopoulos2020influence}. The diffusion process (IC/LT) is introduced for delivering interventions with the FJ opinion dynamics:
\begin{equation}
    x(t+1) = \Lambda W x(t) + (I-\Lambda)s \,,
    \label{eq:FJ}
\end{equation}
where $\Lambda = \mathrm{diag}(\lambda_i)$ encodes node susceptibilities ($\lambda_i \in [0,1]$), $W$ is the influence weight matrix, and $s$ are innate opinions. The steady state is
\[
x^\star(\Lambda) = (I-\Lambda W)^{-1}(I-\Lambda)s \,.
\]

Given a seed set $S \subseteq U$ with $|S| \le k$, diffusion yields an exposure level $\eta_i(S)$ for each node $u_i$. Exposure updates susceptibilities via a decreasing function
\[
\lambda_i' = g(\lambda_i, \eta_i(S)), \quad \frac{\partial g}{\partial \eta} \le 0 \,,
\]
e.g., $g(\lambda,\eta) = \lambda(1-\alpha\eta)$ or $g(\lambda,\eta) = \max\{\lambda - \alpha\eta, \lambda_{\min}\}$. Let $\Lambda' = \mathrm{diag}(\lambda_i')$ and $x^\star(S)$ be the resulting FJ equilibrium.

The influence maximization problem is then:
\[
\max_{S:|S|\le k} \; f_{\text{spread}}(S) \quad \text{s.t.} \quad \mathrm{EC}(x^\star(S)) \ \text{is minimized} \,,
\]
where $f_{\text{spread}}(S)$ is the standard submodular spread objective under IC/LT \citep{kempe2003maximizing} and $\mathrm{EC}(\cdot)$ measures post-intervention echo-chamber intensity \citep{garimella2017reducing}.

\paragraph{\textbf{Diversity-aware seeding across communities}}
Diversity-aware or fairness-aware influence maximization seeks seed sets that not only maximize total spread but also distribute exposure across distinct subpopulations (e.g., communities), so that isolated or minority groups receive meaningful reach, thereby mitigating echo chamber effects and improving cross-cutting exposure. Formalizations typically account for group structure (e.g., detected or given communities) and optimize spread subject to balance or coverage criteria rather than concentrating influence in a few large or central groups \citep{tsang2019group,zhao2024fair,li2020community}.

Let the node set \(U\) be partitioned into communities \(\{C_1,\dots,C_m\}\). Under IC/LT diffusion, let \(\sigma_c(S)\) be the expected number of activations in \(C_c\) from seed set \(S\). A common diversity-aware objective is a concave, weighted coverage:
\[
\max_{S:\,|S|\le k}\;\; \sum_{c=1}^{m} w_c\, f\!\big(\sigma_c(S)\big) \,,
\]
where \(f(\cdot)\) is a concave nondecreasing rewarding balance and \(w_c\!\ge\!0\) encodes group priorities (e.g., to up-weight underserved communities). To avoid size bias, a normalized max-min fairness variant is often used:
\[
\max_{S:\,|S|\le k}\;\; \min_{c}\; \frac{\sigma_c(S)}{|C_c|} \,,
\]
which maximizes the worst-case fraction of communities affected. These formulations encourage the spread of seeds across communities, allowing influence to penetrate otherwise insular groups \citep{tsang2019group,li2020community}.

When the objective is maximizing the influence coverage, the spread function is monotone and submodular, enabling a simple greedy algorithm with a \((1-1/e)\) approximation guarantee \citep{kempe2003maximizing}. Efficient implementations build on this structure, notably CELF for lazy marginal-gain evaluation \citep{leskovec2007cost} and reverse-influence-sampling methods such as TIM and IMM, which achieve near-optimal accuracy with substantially reduced running time \citep{tang2015influence}. In contrast, diversity- or fairness-aware objectives (e.g., max-min coverage across communities, even after normalization by group size) are generally not submodular. Consequently, classic guarantees do not carry over, and one must resort to specialized formulations and approximation schemes developed for fair or community-diversified influence maximization \citep{tsang2019group,li2020community,zhang2019diversifying}. A complementary line of work blends spread with diversity through concave utility aggregation over group-level reach, for which tailored approximation strategies have also been proposed within the influence maximization framework \citep{zhang2019diversifying}.

\paragraph{\textbf{Opinion-aware seeding with stubborn nodes}} The goal is to steer long-run beliefs by placing a small number of stubborn (\textit{zealot}) nodes whose opinions remain fixed, thereby anchoring the dynamics. In opinion-exchange models, introducing fixed-opinion agents can reshape the equilibrium of the remaining population \citep{wu2023identifying}. With appropriate placement and values, this can reduce isolation across communities by weakening within-group reinforcement and injecting bridging signals. This phenomenon is well established for DeGroot/FJ-type processes and for binary variants of the voter model with stubborn nodes. \citep{ghaderi2014opinion,yildiz2013binary}
 
Formally, let \(Z \subseteq U\) be the zealot set with fixed opinions \(o_Z\), and let \(R=U\setminus Z\) denote adaptive nodes. Under a linear update (DeGroot/FJ) with block partitioning of the influence matrix,
\[
x_R^{\star} \;=\; H\,o_Z \;+\; h \,,
\]
where the matrices \(H\) and vector \(h\) depend on network topology and weights (e.g., via the fundamental matrix \((I-W_{RR})^{-1}\)). Thus, the equilibrium on \(R\) is an affine image of \(o_Z\), making both the \emph{locations} and \emph{opinions} of zealots the key control variables. 
 
A design problem of interest is to choose up to \(k\) zealots so as to minimize a post-equilibrium fragmentation measure:
\[
\min_{Z \subseteq V,\; |Z|\le k}\; \mathrm{EC}\!\big(x^{\star}(Z)\big) \,,
\]
where \(\mathrm{EC}(\cdot)\) quantifies echo-chamber intensity (e.g., controversy or cross-community separation). Although this combinatorial objective is generally challenging, several related goals exhibit favorable structure; for example, when the target is to increase mean opinion toward a reference value, the set function over \(Z\) is monotone submodular, enabling greedy selection with approximation guarantees. Empirically and theoretically, the optimized placement of zealots can shift the means and reduce polarization in real networks \citep{hunter2022optimizing}.

\subsection{Content-based Control Mechanisms}
Content-based control steers each user’s exposure toward broader yet still relevant items, anchored in the exposure centroid 
\( \mu_i(t) \) and its dispersion. We organize methods into three families: pre-processing (data and candidate-pool shaping), in-processing (training-time objectives with constraints), and post-processing (re-ranking or allocation with explicit control), following the established taxonomy in Wang et al. \citep{wang2023survey}. These interventions must navigate the accuracy–diversity–utility trade-off, which we address through multi-objective formulations, constrained losses, and constraint-aware re-ranking.

\subsubsection{Pre-processing Approach}

Pre-processing approaches mitigate ideological isolation by directly manipulating input data before model training, making it model-agnostic and relatively easy to integrate \citep{bondielli2019survey}. Most existing works classify pre-processing methods into three categories: \emph{data relabeling}, \emph{re-sampling}, and \emph{data modification} \citep{wang2023survey,li2023fairness}. These strategies address bias in training data distributions, thereby indirectly reducing filter bubbles, selective exposure, and polarization.

\paragraph{\textbf{Data Relabeling}} Data relabeling modifies observed user feedback or labels to correct bias. For example, popularity bias can lead to frequently exposed items receiving disproportionately high labels, thereby reinforcing homogenized recommendations. Relabeling adjusts such labels to reflect underlying user preference rather than exposure frequency \citep{klimashevskaia2024popbias,chen2023bias}. This approach is intuitive and lightweight, but its effectiveness depends on the accuracy of relabeling heuristics.

\paragraph{\textbf{Data Re-sampling}} Resampling methods adjust the distribution of training data by undersampling overrepresented items or oversampling underrepresented content. By balancing exposure frequencies, they enhance diversity and reduce the reinforcement of filter bubbles. For instance, Anand et al. introduce influence-based re-sampling strategies in deep recommender systems to mitigate filter bubbles \citep{anand2022mitigating}, while Abdollahpouri et al. provide a systematic empirical analysis of popularity bias and its connection to fairness and calibration in recommendation \citep{abdollahpouri2020connection}.

\paragraph{\textbf{Data Modification}} Data modification extends beyond resampling and relabeling by actively transforming or augmenting training data. A representative example is counterfactual data augmentation, where new user-item interactions are generated to simulate alternative exposure histories. For instance, Wang et al. propose a sequential recommendation framework that constructs synthetic interaction sequences to increase exposure diversity and mitigate bias arising from imperfect or sparse training data \citep{wang2021counterfactual, huang2023combining}. Recent studies treat data modification as a core pre-processing strategy, noting its role in intervening on interaction or feature representations, but also cautioning about potential distribution shifts or noise introduced by synthetic data when preserving recommendation accuracy \citep{li2023fairness}.

In summary, pre-processing approaches are attractive due to their model-agnostic nature and ease of deployment. Relabeling focuses on correcting biased labels, resampling balances exposure distributions, and modification leverages augmentation or adversarial techniques for stronger debiasing. However, pre-processing alone cannot fully eliminate systemic bias and is often complemented by in-processing and post-processing approaches to comprehensively mitigate ideological isolation.

\subsubsection{In-processing Approach}

In-processing methods embed the mitigation strategy inside the learning or scoring pipeline. Specifically, they modify either the training objective or the scoring function so that the recommender optimizes relevance and mitigation criteria jointly during learning or inference. This makes them expressive, targeting bias, diversity, or fairness, while preserving end-to-end optimization. These methods typically augment the loss or scores as:

\begin{equation}
\mathcal{L} = \mathcal{L}_{rec} + \lambda \cdot \mathcal{L}_{mitigation}, 
\quad 
\hat{y}_{ui} = f_{rec}(u,i) + \Omega(u,i) \,,
\end{equation}
where $\mathcal{L}_{rec}$ is the standard recommendation loss, $\mathcal{L}_{mitigation}$ encodes bias control, fairness constraints, or diversity regularization, and $\lambda>0$ tunes the trade-off. $\hat{y}_{ui}$ represents the \textit{predicted} relevance score for user $u$ on item $i$. The base score $f_{\theta}(u,i)$ models user–item relevance, and $\Omega(u,i)$ is an in-model adjustment, such as debiasing corrections, diversity terms, or counterfactual offsets. We group in-processing into five families as follows.

\paragraph{\textbf{Regularization-based Mitigation}}
Here, explicit penalties or corrective terms are added to reduce exposure/popularity bias during learning. For instance, discrete choice modeling frameworks adjust user-item click probabilities by incorporating exposure frequency:
\begin{equation}
P(i|u, E) = \frac{\exp(\mu_{ui} - \delta \cdot E_i)}{\sum_{j \in  E} \exp(\mu_{uj} - \delta \cdot E_j)} \,,
\end{equation}
thereby attenuating the over-representation of frequently exposed items \citep{krause2024mitigating}.  In addition to the commonly adopted fairness and exposure constraints, Yang et al. introduce content-suppression mechanisms to reduce redundancy and enhance diversity in recommendations \citep{yang2025content}.

\paragraph{\textbf{Latent Representation Modulation}} 
These methods perturb latent representations to encourage diversity along designated semantic directions while preserving relevance. The TD-VAE-CF model achieves this by shifting embeddings along Concept Activation Vectors (CAVs):
\begin{equation}
z'_u = z_u + \alpha \cdot v_{CAV}, \quad \alpha > 0 \,,
\end{equation}
which allows the system to enhance diversity while preserving overall relevance \citep{gao2022mitigating}.

\paragraph{\textbf{Counterfactual Corrections}}
Counterfactual in-processing disentangles causal factors, e.g., preference vs. item popularity, and subtracts the non-preferential component at score time. In MACR, a debiased score removes a direct popularity effect:

\begin{equation}
\hat{y}^{debias}_{ui} = \hat{y}_{ui} - \gamma \cdot \sigma(\hat{y}_i) \,,
\end{equation}
thus reducing systemic popularity bias without altering the user-item matching signal \citep{wei2021model}.

\paragraph{\textbf{Reinforcement Learning (RL)}}
Framing recommendation as an MDP enables temporal exposure control. With state $s_t=(u,H_t)$ (user $u$ and history $H_t$), action $a_t$ (slate or item), and composite reward:

\begin{equation}
R_t = \eta \cdot Acc(a_t) + (1-\eta) \cdot Div(a_t) \,,
\end{equation}
an RL agent learns to trade off short-term accuracy and long-term exposure diversity, mitigating filter-bubble reinforcement \citep{li2023breaking}.

\paragraph{\textbf{Multi-Objective Optimization}}
Multi-objective formulations encode the accuracy-diversity trade-off directly:

\begin{equation}
\max_{\theta} \Big[ \alpha \cdot Acc(\theta) + (1-\alpha) \cdot Div(\theta) \Big] \,,
\end{equation}
where $Acc(\theta)$ measures predictive accuracy, $Div(\theta)$ quantifies exposure diversity, and $\alpha$ tunes the trade-off. Such formulations provide principled ways to reconcile competing objectives in recommender systems \citep{peng2024reconciling}. 

While all in-processing approaches intervene at the model level, their mechanisms differ: regularization-based and counterfactual methods correct statistical biases, latent modulation reshapes semantic representations, RL optimizes sequential decisions, and multi-objective optimization explicitly encodes trade-offs. These approaches demonstrate the versatility of in-processing techniques for increasing fairness and mitigating ideological isolation while retaining relevance.

\subsubsection{Post-processing Approach}

Post-processing methods mitigate ideological isolation by adjusting the items produced by a base recommender model \citep{tang2025model}. Unlike in-processing methods that modify the learning objective or pre-processing approaches that manipulate input data, post-processing provides a flexible, model-agnostic layer for incorporating fairness, diversity, novelty, and neutrality constraints.

Given a candidate set $C$ and target list length $K$, the re-ranked set $R^*(u)$ for user $u$ is defined as:
\begin{equation}
R^*(u) = \arg\max_{S \subseteq C, |S| = K} \; 
\alpha \cdot \underbrace{\sum_{i \in S} f_{rec}(u,i)}_{\text{relevance}} 
+ \beta \cdot \underbrace{D(S)}_{\text{diversity/fairness}} \,,
\end{equation}
where $f_{rec}(u,i)$ measures the relevance of item $i$ for user $u$, 
$D(S)$ quantifies the diversity or fairness of the set $S$, 
and $\alpha, \beta \in [0,1]$ control the trade-off. 
Different methodological schools instantiate this formulation in distinct ways. 

\paragraph{\textbf{Greedy Re-ranking}}

Greedy re-ranking incrementally constructs or revises a top-$K$ list from the base recommender by repeatedly selecting the item that maximizes a weighted combination of relevance and the marginal gain in a set-level criterion, e.g., diversity or fairness. Two classics are MMR ~\citep{carbonell1998use}, which trades off relevance and novelty to reduce redundancy, and xQuAD ~\citep{santos2010exploiting}, which increases coverage over multiple topical aspects during selection. Both are widely used for diversity-aware re-ranking and extend cleanly to recommendation settings. Recent work further enriches the diversity term with LLM-derived semantic signals while keeping the greedy scaffold.
At each step, a typical rule is
\[
i^* = \arg\max_{i \in C \setminus S} \; 
\lambda f_{rec}(u,i) + (1-\lambda) \Delta D(i,S) \,,
\]
where $\Delta D(i,S)$ is the marginal diversity gain of adding $i$ to $S$. 

Recent work explores the use of an LLM as a re-ranker. Specifically, given a candidate list, the LLM is prompted to output a more diverse ranking. Empirically, LLM-based re-ranking improves diversity over random re-ranking but remains below traditional greedy methods (e.g., MMR or xQuAD) on relevance-aware metrics; importantly, the LLM operates as an alternative re-ranker rather than injecting LLM signals into $\Delta D(i,S)$ within a greedy step~\citep{carraro2024enhancing}. Beyond using LLMs as re-rankers, Wang et al. propose a responsible graph-based recommendation pipeline, where a Generative-AI module is applied to synthesize contextualized content based on the user's belief graph and nudge the users toward a more balanced
information perception, mitigating belief filter bubbles ~\citep{wang2024nudging}. 

\paragraph{\textbf{Graph-based and Optimization-based Re-ranking}}
These methods seek a globally balanced set by solving the re-ranking objective over the entire subset $S$ (e.g., $S\subseteq C$, $|S|=K$) rather than making greedy, per-item choices. A common instantiation builds a user-item bipartite graph and imposes flow or coverage constraints to improve catalog balance. For example, FairMatch~\citep{mansoury2020fairmatch} constructs a bipartite graph and uses maximum-flow optimization to boost aggregate diversity by increasing the coverage of underrepresented items. Likewise, Antikacioglu and Ravi minimize discrepancy from a target exposure distribution via minimum-cost flow, achieving explicit control over exposure and coverage ~\citep{antikacioglu2017post}. Compared to greedy heuristics, these flow-based formulations offer stronger fairness guarantees and clearer constraint handling, but their higher computational cost can limit scalability in large, real-time recommender systems.

\paragraph{\textbf{Probabilistic Re-ranking}}
Probabilistic approaches replace an explicit set function $D(S)$ with a likelihood over subsets. A prominent example is the \emph{Determinantal Point Process} (DPP), which defines a distribution $P(S)\propto \det(L_S)$ over item sets $S$, where $L\succeq 0$ is a kernel encoding both item quality and similarity ~\citep{kulesza2012determinantal}. For top-$K$ re-ranking, a common MAP objective is
\[
\max_{S\subseteq C,\;|S|=K}\; \log\det(L_S).
\]
The determinant acts as a repulsive potential: it grows when items are individually high-quality yet mutually dissimilar, thereby discouraging redundancy. A standard parameterization is the quality-similarity factorization
\begin{align}
L &= Q^{1/2} S Q^{1/2}, \\
Q &= \mathrm{diag}(q_1,\dots,q_{|C|}),\quad 
     S_{ii}=1,\quad S_{ij}\in[-1,1] \,,
\end{align}
where $q_i$ is a per-item relevance (quality) score and $S$ captures pairwise similarity. Under this factorization, $\log\det(L_S)$ naturally balances accuracy (via $q_i$) and
diversity (via $S$). Although the exact MAP is generally expensive on large candidate pools, greedy maximization of $\log\det(\cdot)$, a monotone submodular objective, admits an
efficient $(1-1/e)$-approximation with incremental updates,
making DPP re-ranking practical at scale.

\paragraph{\textbf{Novelty-based and Serendipity-based Re-ranking}}
Beyond diversity, post-processing can target \emph{novelty} (items the user has not seen or is unlikely to know) and \emph{serendipity} (items that are both unexpected and useful). A common formalization augments the set objective with per-item novelty or serendipity scores \citep{kotkov2016survey}:
\[
\max_{S\subseteq C,\;|S|=K}\;
\alpha \sum_{i\in S} f_{rec}(u,i)\;+\;
\beta \sum_{i\in S} f_{\text{ser}}(u,i) \,,
\]
where $f_{rec}(u,i)$ denotes base-model relevance and $f_{\text{ser}}(u,i)$ rewards useful surprise. Following prior work, $f_{\text{ser}}$ is often modeled as the product of \emph{usefulness} and \emph{unexpectedness} \citep{xu2020neural}:
\begin{align}
f_{\text{ser}}(u,i) &= \text{Rel}(u,i)\cdot \text{Unexp}(u,i),\\
\text{Unexp}(u,i) &= 1 - \max_{j\in H_u}\!\text{sim}(i,j) \,,
\end{align}
with $H_u$ the user’s history and $\text{sim}(\cdot,\cdot)$ a content or embedding similarity. This construction promotes items that remain relevant while being dissimilar to what the user already knows. In practice, serendipity-oriented re-ranking can improve user satisfaction and long-term engagement, provided the degree of surprise is calibrated. The main design choice is how to instantiate $\text{Unexp}$, e.g., distance to user profile, distance to prior exposures, or popularity-inverse signals, and how
to balance $\alpha$ vs.\ $\beta$ for the target application.

In summary, heuristic approaches provide efficient local approximations, optimization-based methods offer global diversity guarantees, probabilistic approaches deliver principled mathematical formulations, and novelty-oriented methods improve long-term user engagement. All these post-processing techniques can be viewed as instances of the generalized multi-objective framework, differing mainly in how they instantiate the diversity function $D(S)$ and whether they optimize it locally, globally, probabilistically, or with novelty-oriented constraints.

\section{Open Challenges and Future Directions}
\label{sec:challenges_direction}
Having surveyed how ideological isolation can be defined, measured, and mitigated, we now examine what is still required to develop reliable, deployable solutions. This section identifies several cross-cutting open challenges spanning theory, measurement, and system design, and outlines corresponding directions for future research.

 \subsection{Advancing Causal Understanding }

A profound limitation of current quantification metrics is their correlational nature. While such metrics can detect the presence of ideological isolation, they often cannot disentangle its underlying causes \citep{luo2024survey}. For example, a user's ideologically homogeneous feed could arise from either the platform’s personalization algorithm or the user’s selective exposure and homophilic interactions \citep{metzler2024social, milli2025engagement}.

Distinguishing between these drivers is crucial for designing effective interventions. The question here is whether the recommender system should be reconfigured or whether users should be nudged toward more diverse engagement patterns. Future research should adopt causal inference frameworks to distinguish between algorithmic influence and user agency. Approaches such as natural experiments, instrumental variable estimation, and counterfactual modeling, when applied to interaction logs, can help quantify the causal impact of platform-level mechanisms. Moreover, constructing causal models of belief evolution, building on the reinforcement dynamics described in Section~\ref{subsubsec:reinforcement-dynamic}, will be essential for identifying the most effective and ethically appropriate levers for change.

\subsection{Balancing Competing Objectives in Recommender Systems}

A central dilemma in the design of content-based control mechanisms (Section~\ref{subsec:content-metrics}) is the perceived trade-off between recommendation accuracy and ideological diversity \citep{peng2024reconciling,liu2012solving, parapar2021towards}. Many existing in-processing and post-processing approaches risk degrading the perceived utility or relevance of recommendations in pursuit of diversity. If users perceive diversified recommendations as less engaging or misaligned with their interests, they may disengage, undermining the intent to broaden ideological exposure.

A promising research direction is to optimize for a Pareto frontier between accuracy and diversity, rather than treating them as opposing objectives. This can be achieved through adaptive, personalized diversification, in which the degree of exposure to diverse content is dynamically adjusted based on individual user tolerance, susceptibility parameters, and contextual features. In addition, user-controllable interfaces, which allow individuals to calibrate their own diversity-exposure settings, could help reconcile this trade-off by aligning system interventions with user autonomy and engagement preferences.

\subsection{Cross-Platform and Longitudinal Measurement of Ideological Isolation}
\label{subsec:cross_platform_longitudinal}
Another critical challenge is to move beyond single-platform, one-shot studies and actually measure ideological isolation across multiple platforms and over time. In reality, people combine Facebook, X, YouTube, TikTok, news sites, search, and messaging apps into a single "media diet", so isolation is a property of this whole ecosystem, not of Twitter or Facebook alone. Comparative work already shows that echo chambers and polarization appear very different across platforms such as Facebook, Twitter/X, Reddit, and Gab, suggesting that platform-specific snapshots tell only part of the story \citep{cinelli2021echo}. Methodologically, cross-platform identity linkage is difficult, limiting our view of how different forms of isolation co-evolve across a person’s online environment.

Future work should therefore build multiplex and longitudinal measurement frameworks. One direction is to model a user’s environment as a multilayer exposure network, where each layer represents a platform. A second direction is to define isolation trajectories for individuals and groups, tracking how their exposure diversity, cross-cutting contacts, and stance distributions change before, during, and after key events. Finally, agent-based and opinion-dynamics simulations on multiplex networks can serve as testbeds for stress-testing proposed cross-platform metrics under realistic sampling and logging constraints.

\subsection{Robustness to Adversarial and Strategic Amplification}
A key challenge for measuring and mitigating ideological isolation is that the information environment is not purely “organic”: it is actively shaped by adversarial and strategic actors. Social bots, coordinated disinformation campaigns, and manipulation of trending or recommendation systems can all artificially amplify specific narratives, hashtags, or accounts \citep{himelein2021bots}. Recent work shows that relatively small, well-designed campaigns can dramatically increase exposure to harmful or misleading content, exploiting follow networks and recommendation algorithms to maximize coverage \citep{truong2024quantifying}. These operations can deepen echo chambers, manufacture apparent consensus, or crowd out moderate content, thereby distorting any attempt to measure “natural” structural, content-based, or interactional isolation. Yet most existing metrics and models implicitly assume that users, content, and platforms behave non-strategically; they rarely distinguish organic amplification from coordinated or adversarial activity.

Future work needs to treat adversarial and strategic amplification as a first-class concern in the study of ideological isolation. One direction is to design manipulation-aware isolation metrics and measurement pipelines that explicitly detect and discount exposures driven by bots, coordinated campaigns, or inauthentic accounts. Another is to build stress-testing and red-teaming frameworks to see how easily diversity-promoting or depolarization algorithms can be gamed. Finally, robustness methods from recommender systems and graph learning should be directly linked to isolation objectives, so that systems are evaluated not only on accuracy but also on their resistance to being weaponized to create artificial echo chambers.

\section{Conclusion}
\label{sec:conclusion}

This survey has examined ideological isolation in online social networks from a computational perspective, with a particular focus on formal definitions, metrics, computational modeling, and mitigation strategies. 

Our review contributes in three main ways. \textit{First}, we present a unified formal framework that characterizes ideological isolation and its underlying mechanisms, linking user beliefs, content visibility, exposure, and reinforcement dynamics within a shared ideological space. \textit{Second}, we harmonize a multi-view measurement toolkit spanning network-based, content-based, and behavior-based indicators, using consistent notation to support comparability and to clarify apparent discrepancies across metrics. 
\textit{Third}, we organize computational mitigation strategies across modeling and system layers, including Bayesian and agent-based models, LLM-driven simulation, reinforcement learning, and graph neural networks, as well as network-topological and recommendation-pipeline interventions. 

The open challenges identified in this survey point to an inherently interdisciplinary research agenda that integrates network science, machine learning, causal inference, and social science. 
Addressing these challenges calls for responsible, interpretable, and human-centered AI systems that balance personalization with information diversity, thereby fostering healthier digital public spheres and mitigating ideological fragmentation.

\biboptions{numbers,square,sort&compress}
\bibliographystyle{plainnat}   
\bibliography{refs} 

\end{document}